\documentclass[11pt]{article}
\usepackage{amssymb,amscd,array}
\catcode `\@=11
\@addtoreset{equation}{subsection}

\def\harr#1#2{\smash{\mathop{\hbox to .5in{\rightarrowfill}}
\limits^{\scriptstyle#1}_{\scriptstyle#2}}}

\def\harrl#1#2{\smash{\mathop{\hbox to .5in{\leftarrowfill}}
\limits^{\scriptstyle#1}_{\scriptstyle#2}}}

\def\qed{\blacksquare}
\newcommand{\be}{\begin{equation}}
\newcommand{\ee}{\end{equation}}
\newcommand{\bea}{\begin{eqnarray}}
\newcommand{\eea}{\end{eqnarray}}
\newcommand{\R}{\mathbb{R}}
\newcommand{\N}{\mathbb{N}}
\newcommand{\C}{\mathbb{C}}
\newtheorem{thm}{Theorem}[section]
\newtheorem{rem}[thm]{Remark}
\newtheorem{lemma}[thm]{Lemma}
\newtheorem{cor}[thm]{Corollary}
\newtheorem{prop}[thm]{Proposition}
\begin{document}
\begin{titlepage}
\begin{center}
{\bf \Large{Gauge Invariance of the Quantum Electrodynamics ~\\
in the Causal Approach to Renormalization Theory\\}}
\end{center}
\vskip 1.0truecm
\centerline{D. R. Grigore
\footnote{e-mail: grigore@theor1.theory.nipne.ro, grigore@theory.nipne.ro}}
\vskip5mm
\centerline{Dept. of Theor. Phys., Inst. Atomic Phys.}
\centerline{Bucharest-M\u agurele, P. O. Box MG 6, ROM\^ANIA}
\vskip 2cm
\bigskip \nopagebreak
\begin{abstract}
\noindent
We present an extremely simple solution to the renormalization of quantum
electrodynamics based on Epstein-Glaser approach to renormalization theory.
\end{abstract}
\end{titlepage}

\section{Introduction}

The causal approach to renormalization theory pioneered by Epstein and Glaser
\cite{EG1}, \cite{Gl} provides essential simplification at the fundamental
level as well as at to the computational aspects. This is best illustrated in
\cite{Sc1} where quantum electrodynamics is constructed entirely in the
framework of the causal approach. Moreover, one can use the same ideas to
analyse other theories as for instance, Yang-Mills theories \cite{DHKS1}
\cite{DHKS2} \cite{DHS2} \cite{DHS3} \cite{AS} \cite{ASD3} \cite{D1}-\cite{D3}
\cite{Hu1}-\cite{Hu4} \cite {Kr1} \cite{Kr3} \cite{DS}, gravitation 
\cite{Gri1}, \cite{Gri2}, \cite{SW3}, etc.

Let us remind briefly the main ideas of Epstein-Glaser-Scharf approach.
According to Bogoliubov and Shirkov, the $S$-matrix is constructed inductively
order by order as a formal series of operator valued distributions:
\be
S(g)=1+\sum_{n=1}^\infty{i^{n}\over n!}\int_{\R^{4n}} dx_{1}\cdots dx_{n}\,
T_{n}(x_{1},\cdots, x_{n}) g(x_{1})\cdots g(x_{n}),
\label{S}
\ee
where
$g(x)$
is a tempered test function in the Minkowski space 
$\R^{4}$
that switches the interaction and
$T_{n}$
are operator-valued distributions acting in the Fock space of some collection
of free fields. These operator-valued distributions, which are called {\it
chronological products} should verify some properties which can be argued
starting from {\it Bogoliubov axioms}. These axioms will be detailed in the
next Section. The main point is that one can show that, starting from a
convenient {\it interaction Lagrangian}
$T_{1}(x)$
one can construct the whole series
$T_{n}, \quad n \geq 2.$
The interaction Lagrangian must satisfy some requirements such like Poincar\'e
invariance, hermiticity and causality; it is not easy to find a general
solution of this problem but there are some rather general expressions
fulfilling these demands, namely the so-called Wick polynomials. These are
expressions operating  in Hilbert spaces of a special kind, namely in Fock
spaces. A Fock space is a canonical object attached to any single-particle
Hilbert and reasonably describes a system of weakly interacting particles. 

The physical $S$-matrix is obtained from 
$S(g)$ 
taking the {\it adiabatic limit} which is , loosely speaking the limit $ g(x)
\rightarrow 0.$ One should also point out that the recursive process of
constructing the chronological produces fixes them almost uniquely, more
precisely the distribution 
$T_{n}$ 
is unique up to a distribution
$N_{n}(x_{1},\cdots, x_{n})$ 
with support in the set
$$
\{(x_{1},\cdots, x_{n}) \in (\R^{4})^{n} | x_{1}= \cdots = x_{n} \}.
$$
This type of distribution are also called {\it finite renormalizations}. 

In the old version of renormalization theory, one starts from the naive
expressions of the chronological product and sees that they are not properly
defined, i.e. some infinities do appear. The main obstacle is to amend the
naive expression such that well defined expressions are obtained which do also
verify Bogoliubov axioms. In Epstein-Glaser approach, the main problem of the
the construction of the chronological products is done recurringly and it is
reduced to the problem of distribution splitting. It can be proved that this
operation has always solutions consistent with Bogoliubov axioms. 

In the case of a gauge theory there is a supplementary property to be verified.
The main obstacle in constructing the perturbation series for a gauge field
is the fact that, as it happens for the electromagnetic field, one is forced to
use non-physical degrees of freedom for the description of the free fields
\cite{WG}, \cite{SW1}, \cite{Ka} in a Fock space formalism. One must consider
an auxiliary Fock space 
${\cal H}^{gh}$
including, beside the various fields, some fictious fields, called ghosts, and
construct an supercharge that's it an operator $Q$ verifying 
$Q^{2} = 0$ 
such that the physical Hilbert space is 
$
{\cal H}_{phys} \equiv {\it Ker}(Q)/{\it Im}(Q).
$ 
The necessity to consider ghost fields comes mainly from the fact that, up to
now, there is no other way to construct an interaction Lagrangian. On the other
hand, one can construct a convenient interaction Lagrangian in the bigger
Hilbert space
${\cal H}_{gh}$
and apply the construction of Epstein and Glaser without any change. However,
in this case one must impose, beside the usual Bogoliubov axioms, the
supplementary condition that the $S$ matrix factorizes to
${\cal H}_{phys}$.
This condition proves to be too strong and one must replace it by a weaker
condition of factorization to the physical Hilbert space in the adiabatic
limit:
\be
lim_{\epsilon \searrow 0}
\int_{(\R^{4})^{\times n}} dx_{1} \cdots dx_{n}
g(\epsilon x_{1}) \cdots g(\epsilon x_{n})
[Q, T_{n}(x_{1},\dots,x_{n}) ]_{Ker(Q)} = 0,
\quad \forall n \geq 1.
\label{gauge0}
\ee

Even this condition seems to be problematic because the adiabatic limit
does not exists if zero-mass particles are present, so one must weaken further
this requirement as it is done in \cite{DHKS1} where one requires that:
\be
[Q, T_{n}(x_{1},\dots,x_{n})] = i \sum_{l=1}^{n}
{\partial \over \partial x^{\mu}_{l}} 
T^{\mu}_{n/l}(x_{1},\dots,x_{n}),
\quad \forall n \in \N^{*}
\label{gauge1}
\ee
for some Wick polynomials
$T_{n/l}, \quad l = 1, \dots n.$
This condition leads to the previous one if one formally takes the adiabatic
limit and it is called the {\it gauge invariance} of the theory. It is
impressive that this condition for 
$n = 1$
and
$n =2$
fixes almost uniquely the possible form of the interaction Lagrangian and leads
to the presence of a $r$-dimensional Lie group of symmetries in the case of
system composed of $r$ Bosons of spin 1 \cite{AS}, \cite{Gr1}. This result can
be extended to the case of presence of matter fields \cite{ASD3}, \cite{Gr2}
paving the way to a rigorous understanding of the standard model of elementary
particles. In \cite{Gr3} the analysis is pushed to order 
$n = 3$
and the axial anomaly appears in a natural context, as an obstruction to the
factorization condition (\ref{gauge1}) to the physical Hilbert space.  The
gauge invariance problem is now to prove that the identities (\ref{gauge1}) can
be fulfilled for every 
$n \in \N$,
more precisely to show that one can use the freedom left in the chronological
products (the finite renormalizations) to impose gauge invariance in every
order of perturbation theory. This problem is addressed in \cite{Sc1} in the
case of quantum electrodynamics and in \cite{DHKS2} \cite{DHS2} in the
Yang-Mills case. The idea is to assume that one has (\ref{gauge1}) for
$p = 1, \dots, n - 1$
and prove it for
$p = n.$
The proof of renormalizability of quantum electrodynamics from \cite{DHS1},
\cite{DKS2}, \cite{Sc1} is done without using ghosts fields; the same is true
for the proof of the renormalizability of scalar electrodynamics \cite{DKS6}.
This is possible in these two case, but in order to generalize the method to
Yang-Mills theories it is better to understand the proof in the ghost
quantization formalism. Also we mention that the proof from \cite{Sc1} relies
on some reduction formul\ae~ for 1-particle reducible Feynman graphs. A
rigorous proof of these formulae seems to be absent.

The main point of this paper is that with a proper formulation of the induction
hypothesis one can simplify considerably the proof such that it's mathematical
rigor becomes obvious. The idea is that if one formulates the induction
hypothesis in close analogy with the analysis of Epstein and Glaser then one
can prove that in the order $n$ one has instead of (\ref{gauge1}) the relation
\be
[Q, T_{n}(x_{1},\dots,x_{n})] = i \sum_{l=1}^{n}
{\partial \over \partial x^{\mu}_{l}} 
T^{\mu}_{n/l}(x_{1},\dots,x_{n}) + P_{n}(x_{1},\dots,x_{n}),
\quad \forall n \in \N^{*}
\label{anomaly}
\ee
where
$P_{n}$
is a finite renormalization. This finite renormalization can be so much
restricted from the induction process that it is an elementary matter to show
that one can modify appropriately the expressions
$T_{n}$
and
$T_{n/l}$
in such a way that one has 
$P_{n} = 0$.
In this process, the use of some discrete symmetry like charge conjugation is
important. This idea was pioneered in \cite{DHS3} and \cite{D1}.

We will adopt the gauge invariance in the form (\ref{gauge1}) so we will not
touch the adiabatic limit problem in our analysis.

The paper is organized as follows. In the next Section we fix the notations and
clarify the setting we use. We will present Bogoliubov axioms of the
perturbation theory. Because the main point of our paper is to formulate the
induction hypothesis in strict analogy to \cite{EG1} we will summarize the
induction argument used for a theory without gauge invariance. Then we present
the modification of the setting one must impose to study quantum
electrodynamics.  In Section 3 we will give the proof of gauge invariance of
quantum electrodynamics. In Section 4 we present the same analysis for the case
of scalar quantum electrodynamics. The Conclusions are grouped in the last
Section.

\section{Perturbation Theory for QED}

\subsection{Bogoliubov Axioms}{\label{bogoliubov}}

We give here the set of axioms imposed on the chronological products 
$T_{p}$
following the notations of \cite{EG1}. 
\begin{itemize}
\item
First, it is clear that we can consider them {\it completely symmetrical} in
all variables without loosing generality:
\be
T_{p}(x_{\pi(1)},\cdots x_{\pi(p)}) = T_{p}(x_{1},\cdots x_{p}), \quad \forall
\pi \in {\cal P}_{p}.
\label{sym}
\ee
\item
Next, we must have {\it Poincar\'e invariance}. Because we will consider in an
essential way Dirac fields, this amounts to suppose that in the in the Fock
space we have an unitary representation 
$(a, A) \mapsto U_{a, A}$
of the group 
$inSL(2,\C)$
(the universal covering group of the proper orthochronous Poincar\'e group
${\cal P}^{\uparrow}_{+}$ - see \cite{Va} for notations)
such that:
\be
U_{a, A} T_{p}(x_{1},\cdots, x_{p}) U^{-1}_{a, A} =
T_{p}(\delta(A)\cdot x_{1}+a,\cdots, \delta(A)\cdot x_{p}+a), 
\quad \forall A \in SL(2,\C), \forall a \in \R^{4}
\label{invariance}
\ee
where
$SL(2,\C) \ni A \delta(A) \in {\cal P}^{\uparrow}_{+}$
is the covering map.  In particular, {\it translation invariance} is essential
for implementing Epstein-Glaser scheme of renormalization.

Sometimes it is possible to supplement this axiom by corresponding invariance
properties with respect to inversions (spatial and temporal) and charge
conjugation. For the standard model only the PCT invariance is available.
\item
The central axiom seems to be the requirement of {\it causality} which can be
written compactly as follows. Let us firstly introduce some standard notations.
Denote by
$
V^{+} \equiv \{x \in \R^{4} \vert \quad x^{2} > 0, \quad x_{0} > 0\}
$
and
$
V^{-} \equiv \{x \in \R^{4} \vert \quad x^{2} > 0, \quad x_{0} < 0\}
$
the upper (lower) lightcones and by
$\overline{V^{\pm}}$
their closures. If
$X \equiv \{x_{1},\cdots, x_{m}\} \in \R^{4m}$
and
$Y \equiv \{y_{1},\cdots, y_{n}\} \in \R^{4m}$
are such that
$
x_{i} - y_{j} \not\in \overline{V^{-}}, \quad \forall i=1,\dots,m,\quad
j=1,\dots,n
$
we use the notation
$X \geq Y.$
If
$
x_{i} - y_{j} \not\in \overline{V^{+}} \cup \overline{V^{-}}, 
\quad \forall i=1,\dots,m,\quad
j=1,\dots,n
$
we use the notations:
$X \sim Y.$
We use the compact notation
$T(X) \equiv T_{p}(x_{1},\cdots, x_{p})$
with the convention
\be
T(\emptyset) \equiv {\bf 1}
\label{empty}
\ee
and by
$XY$
we mean the juxtaposition of the elements of $X$ and $Y$. In particular,
the expression
$T(X_{1}X_{2})$
makes sense because of the symmetry property (\ref{sym}). Then the causality
axiom writes as follows:
\be
T(X_{1}X_{2}) = T(X_{1}) T(X_{2}), 
\quad \forall X_{1} \geq X_{2}.
\label{causality}
\ee

\begin{rem}
It is important to note that from (\ref{causality}) one can derive easily:
\be
[T(X_{1}), T(X_{2})] = 0, \quad {\rm if}  \quad X_{1} \sim X_{2}.
\label{commute}
\ee
\end{rem}
\item
The {\it unitarity} of the $S$-matrix can be most easily expressed (see
\cite{EG1}) if one introduces, the following formal series:
\be
\bar{S}(g)=1+\sum_{n=1}^\infty{(-i)^{n}\over n!}\int_{\R^{4n}}
dx_{1}\cdots dx_{n}\,\bar{T}_{n}(x_{1},\cdots, x_{n}) g(x_{1})\cdots g(x_{n}),
\label{barS}
\ee
where, by definition:
\be
(-1)^{|X|} \bar{T}(X) \equiv \sum_{r=1}^{|X|} (-1)^{r} \sum_{partitions}
T(X_{1})\cdots T(X_{r});
\label{antichrono}
\ee
here
$X_{1},\cdots,X_{r}$
is a partition of $X$, $|X|$ is the cardinal of the set $X$ and the sum runs
over all partitions. For instance, we have:
\be
\bar{T}_{1}(x) = T_{1}(x)
\ee
and
\be
\bar{T}_{2}(x,y) = - T_{2}(x,y) + T_{1}(x) T_{1}(y) + T_{1}(y) T_{1}(x).
\ee

One calls the operator-valued distributions
$\bar{T_{n}}$
{\it anti-chronological products}. It is not very hard to prove that the series
(\ref{barS}) is the inverse of the series (\ref{S}) i.e. we have:
\be
\bar{S}(g) = S(g)^{-1}
\ee
as formal series. Then the unitarity axiom is:
\be
\bar{T}(X) = T(X)^{\dagger}, \quad \forall X.
\label{unitarity}
\ee

\begin{rem}
One can show that the following relations are identically verified:
\be
\sum (-1)^{|X|} T(X) \bar{T}(Y) = \sum (-1)^{|X|} \bar{T}(X) T(Y) = 0
\label{unit}
\ee
where the sum goes over all partitions 
$X \cup Y = \{1, \dots, p\} \equiv Z, \quad X \cap Y = \emptyset.$
Also one has, similarly to (\ref{causality}):
\be
\bar{T}(X_{1}X_{2}) = \bar{T}(X_{2}) \bar{T}(X_{1}), 
\quad \forall X_{1} \geq X_{2}.
\label{bar-causality}
\ee
\end{rem}
\end{itemize}

A {\it renormalization theory} is the possibility to construct such a
$S$-matrix starting from the first order term:
$
T_{1}(x)
$
which is a Wick polynomial called {\it interaction Lagrangian} which should
verify the following axioms:
\be
U_{a,A} T_{1}(x) U^{-1}_{a,A} = T_{1}(\delta(A)\cdot x+a),
\quad \forall A \in SL(2,\C),
\label{inv1}
\ee
\be
\left[T_{1}(x), T_{1}(y)\right] = 0, \quad \forall x,y \in \R^{4} \quad
s.t. \quad x \sim y,
\label{causality1}
\ee
and
\be
T_{1}(x)^{\dagger} = T_{1}(x).
\label{unitarity1}
\ee

Usually, these requirements are supplemented by covariance with respect to some
discrete symmetries (like spatial and temporal inversions, or PCT), charge
conjugations or global invariance with respect to some Lie group of symmetry.

It is not easy to find non-trivial solutions to the set of requirements
(\ref{inv1}), (\ref{causality1}) and (\ref{unitarity1}). In fact, this is a
problem of constructive field theory. Fortunately, if one considers that the
Hilbert space of the theory is of Fock type, then one has plenty of interesting
solutions, namely the Wick polynomials. As underlined in the Introduction, this
is one of the main reasons of extending the Hilbert space of a gauge system by
including ghost fields: there is no other obvious solution of constructing the
interaction Lagrangian without them.

Let us mention for the sake of the completeness the axioms connected with the
{\it adiabatic limit}, although we will not use them, as we have said in the
Introduction. 

$\bullet$
Let us take in (\ref{S})
$g \rightarrow g_{\epsilon}$
where
$\epsilon \in \R_{+}$
and
\be
g_{\epsilon}(x) \equiv g(\epsilon x).
\ee

Then one requires that the limit
\be
S \equiv \lim_{\epsilon \searrow 0} S(g_{\epsilon})
\label{adiabatic}
\ee
exists, in the weak sense, and is independent of the the test function $g$. In
other words, the operator $S$ should depend only on the {\it coupling constant}
$g \equiv g(0).$
Equivalently, one requires that the limits
\be
T_{n} \equiv \lim_{\epsilon \searrow 0} T_{n}(g_{\epsilon}^{\otimes n}),
\quad n \geq 1
\ee
exists, in the weak sense, and are independent of the test function $g$. One
also calls the limit performed above, the {\it infrared limit}.

$\bullet$
Finally, one demands the {\it stability of the vacuum} and the {\it stability
of the one-particle states} i.e.
\be
\lim_{\epsilon \searrow 0} < \Phi,  S(g_{\epsilon}) \Phi> = 1
\ee
or,
\be
\lim_{\epsilon \searrow 0}
< \Phi,  T_{n}(g^{\otimes n}_{\epsilon}) \Phi> = 0,
\quad \forall n \in \N^{*}
\label{stability}
\ee
if $\Phi$ is the vacuum $\Phi_{0}$ or any one-particle state.

These two requirement amount for the interaction Lagrangian to demand that
\be
T_{1} \equiv \lim_{\epsilon \searrow 0} T_{1}(g_{\epsilon})
\label{adiabatic1}
\ee
should exists, in the weak sense, and should be independent of the test
function $g$. Moreover, we should have
\be
<\Phi, T_{1} \Phi> = 0
\label{stability1}
\ee
if $\Phi$ is the vacuum $\Phi_{0}$ or any one-particle state.

\subsection{Epstein-Glaser Induction}{\label{EG}}

In this Subsection we summarize the steps of the inductive construction of
Epstein and Glaser \cite{EG1}. The main point is a careful formulation of the
induction hypothesis. So, we suppose that we have the interaction Lagrangian 
$T_{1}(x)$
given by a Wick polynomial acting in a certain Fock space. The causality
property (\ref{causality1}) is then automatically fulfilled, but we must make
sure that we also have (\ref{inv1}) and (\ref{unitarity1}).

We suppose that we have constructed the chronological products 
$T_{p}(x_{1},\cdots,x_{p}), \quad p = 1, \dots, n - 1$
having the following properties: (\ref{sym}), (\ref{causality}) and
(\ref{unitarity}) for 
$p \leq n - 1$
and
(\ref{causality}) and (\ref{commute}) for
$|X_{1}| + |X_{2}| \leq n - 1$.
We want to construct the distribution-valued operators
$T(X), \quad |X| = n$
such that the the properties above go from $1$ to $n$.

Here are the main steps of the induction proof.
\begin{enumerate}

\item
One constructs from
$T(X), \quad |X| \leq n - 1$
the expressions
$\bar{T}(X), \quad |X| \leq n - 1$
according to (\ref{antichrono}) and proves the properties
(\ref{bar-causality}) for
$|X| + |Y| \leq n - 1$
and (\ref{unit}) for
$ |Z| \leq n - 1$.

\item
\begin{lemma}
\item
Let us defines the expressions:
\be
A_{n}'(x_{1},\dots,x_{n-1};x_{n}) \equiv {\sum}' (-1)^{|Y|} T(X) \bar{T}(Y),
\ee
\be
R_{n}'(x_{1},\dots,x_{n-1};x_{n}) \equiv {\sum}' (-1)^{|Y|} \bar{T}(X) T(Y) 
\ee
where the sum
${\sum}'$
goes over the partitions 
$
X \cup Y = \{1, \dots, n\}, \quad X \cap Y = \emptyset, \quad 
Y \not= \emptyset, \quad x_{n} \in X.
$

Now, let us suppose that we have a partition
$
P \cup Q = \{1, \dots, n - 1\}, \quad P \cap Q = \emptyset, 
\quad P \not= \emptyset.
$

Then:

If
$Qj \geq P$ one has:
\be
A_{n}'(x_{1},\dots,x_{n-1};x_{n}) = - T(Qn) T(P).
\ee
and if
$Qj \leq P$ one has:
\be
R_{n}'(x_{1},\dots,x_{n-1};x_{n}) = - T(P) T(Qn).
\ee
\label{AR-prime}
\end{lemma}

The proof is elementary if one uses the causality properties (\ref{causality})
and (\ref{bar-causality}).

\item
\begin{cor}
The expression
\be
D_{n}(x_{1},\dots,x_{n-1};x_{n}) \equiv A_{n}'(x_{1},\dots,x_{n-1};x_{n}) -
R_{n}'(x_{1},\dots,x_{n-1};x_{n}).
\label{com-D}
\ee
have causal support i.e.
$supp(D_{n}(x_{1},\dots,x_{n-1};x_{n})) 
\subset \Gamma^{+}(x_{n}) \cup \Gamma^{-}(x_{n})$
where we use standard notations:
\be
\Gamma^{\pm}(x_{n}) \equiv \{ (x_{1},\dots,x_{n}) \in (\R^{4})^{n} |
x_{i} - x_{n} \in V^{\pm} , \quad \forall i = 1, \dots, n-1\}
\ee
\label{D}
\end{cor}

The proof consists of noticing the local character of the support property and 
reducing all possible cases to typical situations from the preceding lemma.

\item
We say that a numerical distribution 
$d(x_{1},\dots,x_{n-1};x_{n})$
is {\it factorizable} (or {\it disconnected}) if it can be written as:
$d(X) = d_{1}(Y) d_{2}(Z)$
where
$Y, Z$ 
is a partition of $X$.

Let us define the {\it canonical dimension} of a Wick monomial 
$\omega(W)$
by assigning to every integer spin field factor and every derivative the value
$1$, for every half-integer spin field factor the value $3/2$ and summing over
all factors.

\begin{lemma}
The distribution 
$D_{n}(x_{1},\dots,x_{n-1};x_{n})$
can be written as a sum
\be
D_{n}(x_{1},\dots,x_{n-1};x_{n}) = \sum_{i} d_{i}(x_{1},\dots,x_{n-1};x_{n})
W_{i}(x_{1},\dots,x_{n-1};x_{n})
\ee
where
$W_{i}(x_{1},\dots,x_{n-1};x_{n})$
are linearly independent Wick monomials and
$d_{i}(x_{1},\dots,x_{n-1};x_{n})$
are numerical distributions with causal support i.e
$
supp(d_{i}(x_{1},\dots,x_{n-1};x_{n})) 
\subset \Gamma^{+}(x_{n}) \cup \Gamma^{-}(x_{n}).
$
Moreover, the set of Wick monomials appearing in the preceding formula can be
obtained from the expression
$
T_{1}(x_{1}) \cdots T_{1}(x_{n})
$
by taking all possible Wick contractions, eliminating the monomials for for
which the corresponding numerical distributions are factorizable and keeping a
linearly independent set.

Finally, the following limitations are valid:
\be
\omega(d_{i}) + \omega(W_{i}) \leq \omega(T_{1}), \quad \forall i.
\label{deg-d}
\ee
\label{wick}
\end{lemma}
The proof goes by induction. 

\item
If
$d = (d_{i})_{i = 1}^{N}$
is a multi-component distribution and
$SL(2,\C) \ni A \rightarrow D(A)$
is an $N$-dimensional representation of the group
$SL(2,\C)$
we define a new distribution according to:
\be
(A \cdot d)(x) \equiv D(A) d (\delta(A^{-1}) \cdot x)
\ee
and say that the distribution $d$ is
$SL(2,\C)$-{\it covariant} 
{\it iff} it verifies:
\be
A \cdot d = d.
\ee
We remark that we have defined a
$SL(2,\C)$ 
action:
\be
(A_{1} \cdot A_{2}) \cdot d = A_{1} \cdot (A_{2} \cdot d), \quad 
{\bf 1} \cdot d = d.
\ee
For such multi-component distribution, the order of singularity 
$\omega(d)$
is, by the definition, the maximum of the orders of singularities of the
components.
\begin{lemma}
The distributions
$d_{i}(x_{1},\dots,x_{n-1};x_{n})$
defined above are 
$SL(2,\C)$-covariant.
\end{lemma}

The proof follows from the induction hypothesis (\ref{invariance}).
\item
Now we have the following result from \cite{DHKS2}, \cite{Sc1}:
\begin{lemma}
Let $d$ be a
$SL(2,\C)$-covariant
distribution with causal support. Then, there exists a causal splitting
\be
d = a - r, \quad supp(a) \subset \Gamma^{+}(x_{n}), \quad 
supp(r) \subset \Gamma^{-}(x_{n})
\ee
which is also
$SL(2,\C)$-covariant and such that 
\be
\omega(a) \leq \omega(d), \quad \omega(r) \leq \omega(d).
\ee
\label{cov-split}
\end{lemma}
We outline the proof because the argument is generic and it will also be used 
for the more general case of gauge invariance. It is known from the general 
theory of distribution splitting that there exists a causal splitting
$d = a - r$
preserving the order of singularity. Then
$A \cdot d = A \cdot a - A \cdot r$
is a causal splitting of the distribution 
$A \cdot d$.
Because, by hypothesis, we have
$A \cdot d = d$
it follows that we have
\be 
A \cdot a -a = A \cdot r - r.
\label{cs}
\ee
But the left hand side has support in
$\Gamma^{+}(x_{n})$
and the right hand side in
$\Gamma^{-}(x_{n})$
so, the common value, denoted by 
$P_{A}$
have the support in 
$
\Gamma^{+}(x_{n}) \cap \Gamma^{-}(x_{n}) = 
\{(x_{1},\cdots, x_{n}) \in (\R^{4})^{n} | x_{1}= \cdots = x_{n} \}.
$
But in this case, it is known from the general distribution theory that 
$P_{A}$
is of the form
\be
P_{A}(x) = p(\partial) \delta^{n-1}(X)
\ee
where
\be
\delta^{n-1}(X) \equiv \delta(x_{1}-x_{n}) \cdots \delta(x_{n-1}-x_{n})
\ee
and $p$ is a polynomial in the derivatives of maximal order 
$\omega(d)$.
In particular, if
$\omega(d) < 0$
we have 
$p = 0$
and the causal splitting is 
$SL(2,\C)$-covariant.
If
$\omega(d) \geq 0$
then we easily derive that
$P_{A}$
verifies the following identity:
\be
P_{A_{1}\cdot A_{2}} = P_{A_{1}} + A_{1} \cdot P_{A_{2}}.
\ee
This relation says that the map
$A \rightarrow P_{A}$
is a
$SL(2,\C)$-cocycle 
with values in the finite dimensional space of polynomials of order not greater
than 
$\omega(d)$.
Because
$SL(2,\C)$
is a connected, simply connected and simple Lie group we can apply Hochschild 
lemma \cite{Va} and obtain that
$P_{A}$
is of the form
\be
P_{A} = A_{1} \cdot Q - Q
\ee
for some polynomial $Q$ of order not greater than 
$\omega(d)$.
In particular, we have
\be
A \cdot (a - Q) = a - Q
\ee
so we have a
$SL(2,\C)$-covariant 
causal splitting
$d = (a - Q) - (r -Q)$.
$\qed$

\item
\begin{cor}
There exists a 
$SL(2,\C)$-covariant causal splitting:
\be
D_{n}(x_{1},\dots,x_{n-1};x_{n}) =
A_{n}(x_{1},\dots,x_{n-1};x_{n}) - R_{n}(x_{1},\dots,x_{n-1};x_{n})
\label{decD}
\ee
with
$supp(A_{n}(x_{1},\dots,x_{n-1};x_{n})) \subset \Gamma^{+}(x_{n})$
and
$supp(R_{n}(x_{1},\dots,x_{n-1};x_{n})) \subset \Gamma^{-}(x_{n})$.
\end{cor}
For that reason, the expressions
$A_{n}$ 
and
$R_{n}$
are called {\it advanced} (resp. {\it retarded}) products.
\item
\begin{lemma}
The following relation is true
\be
D_{n}(x_{1},\dots,x_{n-1};x_{n})^{\dagger} = (-1)^{n-1}
D_{n}(x_{1},\dots,x_{n-1};x_{n}).
\ee
In particular the causal splitting obtained above can be chosen such that
\be
A_{n}(x_{1},\dots,x_{n-1};x_{n})^{\dagger} = (-1)^{n-1}
A_{n}(x_{1},\dots,x_{n-1};x_{n}).
\ee
\label{unitD}
\end{lemma}
The first assertion follows by elementary computations starting directly from
the definition (\ref{com-D}) and using the unitarity induction hypothesis
(\ref{unitarity}) and the relations (\ref{unit}). This proves that by
performing the substitutions:
\bea
A_{n}(x_{1},\dots,x_{n-1};x_{n}) \rightarrow {1\over 2}
\left[ A_{n}(x_{1},\dots,x_{n-1};x_{n})^{\dagger} + (-1)^{n-1}
A_{n}(x_{1},\dots,x_{n-1};x_{n})\right]
\nonumber \\
R_{n}(x_{1},\dots,x_{n-1};x_{n}) \rightarrow {1\over 2}
\left[ R_{n}(x_{1},\dots,x_{n-1};x_{n})^{\dagger} + (-1)^{n-1}
R_{n}(x_{1},\dots,x_{n-1};x_{n})\right] 
\eea
we do not affect the relation from the preceding corollary and we obtain a
causal splitting verifying the relation from the statement without spoiling the
$SL(2,\C)$-covariance.
$\qed$

\item
Now we have 
\begin{thm}
Let us define
\bea
T_{n}(x_{1},\cdots, x_{n}) \equiv 
A_{n}(x_{1},\cdots, x_{n-1};x_{n}) - A_{n}'(x_{1},\cdots, x_{n-1};x_{n}) 
\nonumber \\ \equiv 
R_{n}(x_{1},\cdots, x_{n-1};x_{n}) - R_{n}'(x_{1},\cdots, x_{n-1};x_{n}).
\eea
Then these expressions satisfy the 
$SL(2,\C)$-covariance, 
causality and unitarity conditions (\ref{invariance}) (\ref{causality}) 
(\ref{commute}) and (\ref{unitarity}) for
$p = n$.
If we substitute
\be
T_{n}(x_{1},\cdots, x_{n}) \rightarrow {1 \over n!}
\sum_{\pi} T_{n}(x_{\pi(1)},\cdots, x_{\pi(n)})
\ee
where the sum runs over all permutations of the numbers
$\{1, \dots, n\}$
then we also have the symmetry axiom (\ref{sym}). The generic expression of the
chronological product is similar to that appearing in lemma \ref{wick} 
\be
T_{n}(x_{1},\dots,x_{n}) = \sum_{i} t_{i}(x_{1},\dots,x_{n-1};x_{n})
W_{i}(x_{1},\dots,x_{n-1};x_{n})
\label{tn}
\ee
with the same limitation (\ref{deg-d}) on the numerical distributions:
\be
\omega(t_{i}) + \omega(W_{i}) \leq \omega(T_{1}), \quad \forall i.
\label{deg-t}
\ee
\label{chronos}
\end{thm}
The 
$SL(2,\C)$-covariance 
is obvious. The causality axiom (\ref{causality}) follows from the two
expressions of the definition of
$T_{n}$
if one takes into account the support properties of the advanced and retarded
product and also uses lemma \ref{AR-prime}.. The property (\ref{commute})
follows from general properties of the Wick monomials. The unitarity axiom is a
result of the definition given above, the property of the advanced products
from the preceding lemma, the expressions
$A_{n}'$
and the induction hypothesis (\ref{unitarity}) for 
$p \leq n - 1$. 
The symmetrization process is obvious.
$\qed$
\end{enumerate}

As we have mentioned in the Introduction the solution of the renormalization
problem is not unique. The non-uniqueness is given by the possibility of adding
to the distributions
$T_{n}$
some finite renormalizations
$N_{n}$.
There are some restrictions on these finite renormalizations coming from the
Poincar\'e invariance and unitarity but still there remains some arbitrariness.
One can restrict even further the arbitrariness requiring the existence of the
adiabatic limit. One can prove that this limit does exists if there are no
zero-mass particles in the spectrum of the energy-momentum quadri-vector.

\subsection{Perturbation Theory for Zero-Mass Particles}{\label{zero}}

We remind the basic facts about the quantization of the photon; for more
details see \cite{Gr1} and references quoted there.  Let us denote the Hilbert
space of the photon by
${\rm H}_{photon}$;
it carries the unitary representation of the orthochronous Poincar\'e group 
${\sf H}^{[0,1]} \oplus {\sf H}^{[0,-1]}$
(see \cite{Va}).

The Hilbert space of the multi-photon system should be, according to the basic
principles of the second quantization, the associated symmetric Fock space
${\cal F}_{photon} \equiv {\cal F}^{+}({\rm H}_{photon})$.
One can construct in a rather convenient way this Fock space in the spirit of
algebraic quantum field theory. One considers the Hilbert space 
${\cal H}^{gh}$
generated by applying on the vacuum 
$\Phi_{0}$ 
the free fields 
$A^{\mu}(x), \quad u(x), \quad \tilde{u}(x)$ 
called the {\it electromagnetic potential} (reps. {\it ghosts}) which are
completely characterize by the following properties:

\begin{itemize}
\item
Equation of motion:
\be
\square A^{\mu}(x) = 0, \quad \square u(x) = 0, \quad \square \tilde{u}(x) = 0.
\label{equ}
\ee
\item
Canonical (anti)commutation relations: 
\bea
[A^{\mu}(x),A^{\rho}(y)] = -g^{\mu\rho} D_{0}(x-y) \times {\bf 1}, 
\quad [A^{\mu}(x),u(y)] = 0, \quad [A^{\mu}(x),\tilde{u}(y)] = 0
\nonumber \\
\{u(x),u(y)\} = 0, \quad \{\tilde{u}(x),\tilde{u}(y)\} = 0, \quad
\{u(x),\tilde{u}(y)\} = D_{0}(x-y) \times {\bf 1};
\label{CCR}
\eea
here 
$D_{m}, \quad m \geq 0$
is the Pauli-Jordan distribution:
\be
D_{m}(x) \equiv {1 \over (2\pi)^{3}} \int_{\R^{3}} 
{d{\bf p}  \over 2 \sqrt{{\bf p}^{2} + m^{2}}} 
exp(-i x_{0} \sqrt{{\bf p}^{2} + m^{2}} + i {\bf x} \cdot {\bf p}).
\label{PJ}
\ee
\item
Covariance properties with respect to the Poincar\'e group. Let
$I_{s}$
and
$I_{t}$
be the space (time) inversion in the Minkowski space 
$\R^{4}$.
Let
$U_{a,A}, \quad U_{I_{s}}$
be the unitary operators realizing the 
$SL(2,\C)$
transformations and the spatial inversion respectively and
$U_{I_{t}}$
the anti-unitary operator realizing the temporal inversion; then we require:
\bea
U_{a,A} A^{\mu}(x) U^{-1}_{a,A} = {\delta(A^{-1})^{\mu}}_{\nu}
A^{\nu}(\delta(A) \cdot x + a), 
\nonumber \\
U_{a,A} u(x) U^{-1}_{a,A} = u(\delta(A) \cdot x + a),
\quad
U_{a,A} \tilde{u}(x) U^{-1}_{a,A} = \tilde{u}(\delta(A) \cdot x + a)
\label{poincare}
\eea
\bea
U_{I_{s}} A^{\mu}(x) U_{I_{s}}^{-1} = 
{(I_{t})^{\mu}}_{\nu} A^{\nu}(I_{s}\cdot x),
\quad
U_{I_{s}} u(x) U_{I_{s}}^{-1} = - u(I_{s}\cdot x),
\quad
U_{I_{s}} \tilde{u}(x) U_{I_{s}}^{-1} = - \tilde{u}(I_{s}\cdot x);\quad
\label{spatial}
\eea
\bea
U_{I_{t}} A^{\mu}(x) U_{I_{t}}^{-1} = 
{(I_{t})^{\mu}}_{\nu} A^{\nu}(I_{t}\cdot x),
\quad
U_{I_{t}} u(x) U_{I_{t}}^{-1} = - u(I_{t}\cdot x),
\quad
U_{I_{t}} \tilde{u}(x) U_{I_{t}}^{-1} = - \tilde{u}(I_{t}\cdot x).\quad
\label{temporal}
\eea
The spatio-temporal inversion is:
$U_{I_{st}} \equiv U_{I_{s}}~U_{I_{t}}$.
\item
Charge invariance. The unitary operator realizing the charge conjugation
verifies: 
\bea
U_{C} A^{\mu}(x) U_{C}^{-1} = - A^{\mu}(x),
\quad
U_{C} u(x) U_{C}^{-1} = - u(x),
\quad
U_{C} \tilde{u}(x) U_{C}^{-1} = - \tilde{u}(x).
\label{charge}
\eea
\item
Moreover, we suppose that these operators are leaving the vacuum invariant:
\be
U_{a,A} \Phi_{0} = \Phi_{0}, \quad 
U_{I_{s}} \Phi_{0} = \Phi_{0}, \quad 
U_{I_{t}} \Phi_{0} = \Phi_{0}, \quad 
U_{C} \Phi_{0} = \Phi_{0}.
\label{inv-vacuum}
\ee
\end{itemize}

\begin{rem}
One can easily prove that the operators
$U_{a,A}, \quad U_{I_{s}}$
and
$U_{I_{t}}$
are realizing a projective representation of the Poincar\'e group i.e. they
have suitable commutation properties (see \cite{Va} rel. (196) from ch. IX. 6).
Also the charge conjugation operator commutes with these operators. (As it is
well known, there is some freedom in choosing some phases in the definitions of
the spatial and temporal inversions \cite{Va}; we have made the convenient
choice which ensures this commutativity property).
\label{projective}
\end{rem}

\begin{rem}
One can prove that all the operators 
$U_{\dots}$
defined above are leaving the commutation relations invariant. This fact can be
used to prove that they are unitary (anti-unitary).
\end{rem}
We suppose that in
${\cal H}^{gh}$
we have, beside the scalar product, a sesqui-linear form
$<\cdot,\cdot>$
and we denote the conjugate of the operator $O$ with respect to this form by 
$O^{\dagger}$.
One can completely characterize this form by requiring:
\be
A_{\mu}(x)^{\dagger} = A_{\mu}(x), \quad
u(x)^{\dagger} = u(x), \quad
\tilde{u}(x)^{\dagger} = - \tilde{u}(x).
\label{conjugate}
\ee

Now, we define in
${\cal H}^{gh}$
an important operator called {\it supercharge} according to:
\be
Q = \int_{\R^{3}} d^{3}x  \partial^{\mu} A_{\mu}(x)
\stackrel{\leftrightarrow}{\partial_{0}}u(x)
\label{supercharge}
\ee
and one can prove the following properties:
\be
Q \Phi_{0} = 0
\label{Q-0}
\ee
and
\be
\{Q,u(x)\} = 0,\quad
\{Q,\tilde{u}(x)\} =  - i \partial^{\mu} A_{\mu}(x),\quad
[Q, A_{\mu}(x)] = i \partial_{\mu} u(x).
\label{Q-com}
\ee
From these properties one can derive
\be
Q^{2} = 0;
\label{square}
\ee
so we also have
\be
Im(Q) \subset Ker(Q).
\label{im-ker}
\ee

Next, we denote by ${\cal W}$ the linear space of all Wick monomials acting in 
the Fock space
${\cal H}^{gh}$
generated by the fields
$A_{\mu}(x),~ u(x)$
and
$\tilde{u}(x)$.
\begin{rem}
We notice that usually one constructs Wick monomials by first decomposing every
the free fields in a creation and an annihilation parts and then ordering the
creation parts to the left with respect to the annihilation parts (with the
corresponding Jordan sign if Fermion fields are present). However, there is a
way to define Wick monomials without this decomposition, by a substraction
procedure \cite{SW2} pg. 104; for instance:
\be
:A^{\mu}(x) A^{\nu}(x): \equiv lim_{x_{1},x_{2} \rightarrow x}
[A^{\mu}(x_{1}) A^{\nu}(x_{2}) - 
(\Phi_{0},A^{\mu}(x_{1}) A^{\nu}(x_{2})\Phi_{0})]
\ee
\end{rem}

If $M$ is such a Wick monomial, we define by
$gh_{\pm}(M)$
the degree in $u$ (resp. in $\tilde{u}$). The {\it ghost number} is, by
definition, the expression:
\be
gh(M) \equiv gh_{+}(M) - gh_{-}(M).
\ee
Then we  define the operator:
\be
d_{Q} M \equiv :QM: - (-1)^{gh(M)} :MQ:
\label{BRST-op}
\ee
on monomials $M$ and extend it by linearity to the whole 
${\cal W}$. 
The operator
$d_{Q}: {\cal W} \rightarrow {\cal W}$
is called the {\it BRST operator}; its properties are following elementary from
the properties of the supercharge: beside the Leibnitz rule we have:
\be
d_{Q} u = 0, \quad d_{Q} \tilde{u} = - i \partial^{\mu} A_{\mu}, \quad
d_{Q} A_{\mu} = i \partial_{\mu} u.
\label{BRST}
\ee
As a consequence of (\ref{square}), it verifies:
\be
d_{Q}^{2} = 0.
\label{Q2}
\ee

Now one can prove that for any Wick monomial $W$ we have:
\be
U_{g} (d_{Q} W) U_{g}^{-1} = d_{Q} U_{g} W U_{g}^{-1},
\quad \forall g = (a,A), I_{s}, I_{t}, C.
\label{uw}
\ee
The proof of this relation is elementary if $W$ is one of the basic fields
$A^{\mu}, \quad u, \quad \tilde{u}$. Then one extends it by induction for a
Wick monomial with an arbitrary number of factors. As a corollary we have:
\be
U_{g} Q = Q U_{g}, \quad \forall g = (a,A), I_{s}, I_{t}, C.
\label{UQ}
\ee

Then we have the central result
\begin{thm}
The sesqui-linear form 
$<\cdot,\cdot>$
factorizes to a well-defined scalar product on the completion of the factor
space 
$Ker(Q)/Im(Q)$.
Then there exists the following Hilbert spaces isomorphism:
\be
\overline{Ker(Q)/Im(Q)} \simeq {\cal F}_{photon};
\label{factor}
\ee
The representation of the Poincar\'e group and the charge conjugation operator 
are factorizing to 
$Ker(Q)/Im(Q)$
and are producing unitary operators with the exception of the temporal (and
spatio-temporal) inversions which are anti-unitary.
\label{photon+ghosts}
\end{thm}

We will need the following relation in the next Section:
\be
(d_{Q}W)^{\dagger} = - (-1)^{gh(W)} d_{Q} (W^{\dagger})
\label{dag-w}
\ee
valid for any Wick monomial. The proof is similar to the proof of (\ref{uw}):
one shows elementary that the relation is true for any of the basic fields
$A^{\mu}, \quad u, \quad \tilde{u}$
and extends it to an arbitrary product by induction.

We remind that if
$O$
is a self-adjoint operator verifying the condition
\be
d_{Q} O = 0
\label{dQ}
\ee
then it induces a well defined operator
$[O]$
on the factor space
$\overline{Ker(Q)/Im(Q)} \simeq {\cal F}_{photon}$.
This kind of observables on the physical space are called {\it gauge
invariant observables}. However, the operators of the type
$d_{Q} O$
are inducing a null operator on the factor space, so are not interesting.
For more general considerations on this point see \cite{DF}.

Usually one has to add into the game {\it matter} fields. These are operators
for which one has to give separately the corresponding canonical
(anti)commutation relations and transformation rules with respect to the
Poincar\'e group and charge conjugation. By definition, we keep the same
expression for the supercharge and construct the physical Hilbert space by the
same factorization procedure. In particular, this will mean that the BRST
operator acts trivially on the matter fields. 

We can formulate now what we mean by a perturbation theory of electromagnetism
+ matter. By definition, this means that we can construct in
${\cal H}^{gh}$
the set of chronological products
$T_{n}$
as in the Subsection \ref{bogoliubov} and we impose in addition a factorization
condition to the physical Hilbert space. To avoid infra-red divergence
problems, we adopt as said in the Introduction the condition (\ref{gauge1})
which we prefer to write into the form:
\be
d_{Q} T_{n}(x_{1},\dots,x_{n}) = i \sum_{l=1}^{n}
{\partial \over \partial x^{\mu}_{l}} 
T^{\mu}_{n/l}(x_{1},\dots,x_{n}),
\quad \forall n \in \N^{*}
\label{gauge2}
\ee
for some Wick polynomials
$T_{n/l}, \quad l = 1, \dots n.$

By definition, this is the {\it gauge invariance} condition. It can be
connected with the usual approaches based on the Ward identities imposed on
the (renormalized) Feynman distributions.

Let us note that the Wick polynomials
$T_{n/l}, \quad l = 1, \dots n$,
if they exists, are highly non-unique.

Usually, one can extend the definition of the various symmetries of the theory
(Poincar\'e covariance and charge conjugation invariance) to the case of
electromagnetism + matter in such a way that all the properties mentioned
before remain true.

\section{Renormalizability of Quantum Electrodynamics}{\label{qed}}

\subsection{The Interaction Lagrangian}{\label{int-qed}}

By definition, in this case the matter field is a Dirac field of mass $m$
denoted by
$\psi(x) = \psi_{\alpha}(x)_{\alpha=1}^{4}$.
To describe this field we need Dirac matrices 
$\gamma^{\mu}, \quad \mu = 0,\dots,3$
for which we prefer the chiral representation \cite{Va}:
\be
\gamma_{0} = \left( \matrix{ 0 & 1 \cr 1 & 0} \right), \quad
\gamma_{i} = \left( \matrix{ 0 & - \sigma_{i} \cr \sigma_{i} & 0} \right),
\quad i = 1,2,3;
\ee
here
$\sigma_{i}, \quad i = 1,2,3$
are the Pauli matrices. This is a representations in which the matrix
$\gamma_{5} \equiv i \gamma_{0}\gamma_{1}\gamma_{2}\gamma_{3}$
is diagonal:
\be
\gamma_{5} = \left( \matrix{ 1 & 0 \cr 0 & -1} \right).
\ee

We denote as usual
$\bar{\psi}(x) \equiv \psi(x)^{*} \gamma_{0}$;
it is convenient to consider $\psi$ ($\bar{\psi}$) as a column (line) vector.
As before, the Dirac field is characterized by:

\begin{itemize}
\item
Equation of motion (which is, of course the {\it Dirac equation}):
\be
(i \gamma \cdot \partial + m) \psi(x) = 0.
\label{dirac-equ}
\ee
\item
Canonical (anti)commutation relations: 
\bea
[\psi(x),A^{\mu}(y)] = 0, \quad [\psi(x),u(y)] = 0, 
\quad [\psi(x),\tilde{u}(y)] = 0
\nonumber \\
\{\psi(x),\psi(y)\} = 0, \quad 
\{\psi(x),\bar{\psi}(y)\} = S_{m}(x-y) \times {\bf 1};
\label{CAR}
\eea
here 
$S_{m}, \quad m \geq 0$
is a $4 \times 4$ matrix given by:
\be
S_{m}(x) \equiv (i \gamma \cdot \partial + m) D_{m}(x).
\label{Sm}
\ee
\item
Covariance properties with respect to the Poincar\'e group:
\bea
U_{a,A} \psi(x) U^{-1}_{a,A} = S(A^{-1}) \psi(\delta(A) \cdot x + a), 
\nonumber \\
U_{I_{s}} \psi(x) U_{I_{s}}^{-1} = i \gamma_{0} \psi(I_{s}\cdot x),
\quad
U_{I_{t}} \psi(x) U_{I_{t}}^{-1} = -i \gamma_{1} \gamma_{3} \psi(I_{s}\cdot x);
\label{dirac-cov}
\eea
here 
\be
S(A) \equiv \left( \matrix{ A & 0 \cr 0 & (A^{-1})^{*}} \right).
\ee
\item
Charge invariance. The unitary operator realizing the charge conjugation
verifies: 
\be
U_{C} \psi(x) U_{C}^{-1} = \gamma_{0} \gamma_{2} \bar{\psi}(x)^{t}.
\label{dirac-charge}
\ee
\end{itemize}

These relations should be added to the ones from the preceding Subsection. It
can be proved that Remark \ref{projective} stays true. 

By definition, the interaction Lagrangian is:
\be
T_{1}(x) \equiv e :\bar{\psi}(x) \gamma_{\mu} \psi(x): A^{\mu}(x)
\label{T1}
\ee
(here $e$ is a real constant the electron charge) and one can verify easily
that the properties (\ref{inv1}), (\ref{causality1}) and (\ref{unitarity1}) are
true. Moreover, we have 
\be
U_{I_{s}} T_{1}(x) U^{-1}_{I_{s}} = T_{1}(I_{s}\cdot x),\quad
U_{I_{t}} T_{1}(x) U^{-1}_{I_{t}} = T_{1}(I_{t}\cdot x),\quad
U_{C} T_{1}(x) U^{-1}_{C} = T_{1}(x).
\label{pct-qed}
\ee

The most important property is (\ref{gauge2}) for 
$n = 1$ 
with
\be
T_{1/1}^{\mu}(x) \equiv e :\bar{\psi}(x) \gamma^{\mu} \psi(x): u(x).
\label{T1/1}
\ee

We list below some obvious properties of the preceding expressions which are
similar to the properties (\ref{inv1}), (\ref{causality1}), (\ref{unitarity1})
and (\ref{pct-qed}):
\bea
U_{a,A} T^{\mu}_{1/1}(x) U^{-1}_{a,A} = {\delta(A^{-1})^{\mu}}_{\rho}
T^{\rho}_{1/1}(\delta(A)\cdot x+a), \quad \forall A \in SL(2,\C),
\nonumber \\
U_{I_{s}} T^{\mu}_{1/1}(x) U^{-1}_{I_{s}} = {(I_{s})^{\mu}}_{\rho}
T^{\rho}_{1/1}(I_{s}\cdot x),\quad
U_{I_{t}} T^{\mu}_{1/1}(x) U^{-1}_{I_{t}} = {(I_{s})^{\mu}}_{\rho}
T^{\rho}_{1/1}(I_{t}\cdot x), \quad
\nonumber \\
U_{C} T^{\mu}_{1/1}(x) U^{-1}_{C} = T^{\mu}_{1/1}(x).
\label{inv1/1}
\eea
\be
\left[T^{\mu}_{1/1}(x), T^{\rho}_{1/1}(y)\right] = 0, \quad 
\left[T^{\mu}_{1/1}(x), T_{1}(y)\right] = 0, \quad 
\forall x,y \in \R^{4} \quad s.t. \quad x \sim y,
\label{causality1/1}
\ee
\be
T^{\mu}_{1/1}(x)^{\dagger} = T^{\mu}_{1/1}(x)
\label{unitarity1/1}
\ee
and
\be
gh(T_{1}) = 0, \quad gh(T^{\mu}_{1}) = 1.
\label{gh1/1}
\ee

These properties can be easily deduced from the definitions of the various
symmetry transformations and the explicit expression (\ref{T1/1}).

From the explicit expressions of the second order chronological product
obtained in \cite{Sc1} one can verify that we have similar relations in the
second order of perturbation theory.

We close by mentioning that properties (\ref{uw}), (\ref{UQ}) and (\ref{dag-w})
remain true.

\subsection{Gauge Invariance of Quantum Electrodynamics}{\label{ren-qed}}

In this Subsection we prove the following theorem.
\begin{thm}
Suppose that
$T_{1}$
is given by (\ref{T1}) above. Then the distributions
$T_{n}$
can be constructed such that, beside the conditions of symmetry,
$SL(2,\C)$-covariance,
causality and unitarity (\ref{sym}), (\ref{invariance}), (\ref{causality}), 
(\ref{unitarity}), also verify covariance with respect to spatial and temporal
inversions, invariance to charge conjugation and gauge invariance:
\be
U_{I_{s}} T_{n}(x_{1},\dots,x_{n}) U^{-1}_{I_{s}} = 
T_{n}(I_{s}\cdot x_{1},\dots,I_{s}\cdot x_{n}),
\label{spatial-n}
\ee
\be
U_{I_{t}} T_{n}(x_{1},\dots,x_{n}) U^{-1}_{I_{t}} = 
\bar{T}_{n}(I_{t}\cdot x_{1},\dots,I_{t}\cdot x_{n}),
\label{temporal-n}
\ee
\be
U_{C} T_{n}(x_{1},\dots,x_{n}) U^{-1}_{C} =  T_{n}(x_{1},\dots,x_{n}),
\label{charge-n}
\ee
\be
d_{Q} T_{n}(x_{1},\dots,x_{n}) =  i \sum_{l=1}^{n}
{\partial \over \partial x^{\mu}_{l}} 
T^{\mu}_{n/l}(x_{1},\dots,x_{n}),
\quad \forall n \in \N^{*}
\label{gauge-n}
\ee
where
$T^{\mu}_{n/l}, \quad l = 1,\dots,n$
are some Wick monomials.
\label{gaugeQED}
\end{thm}
{\bf Proof:}

(i) The main trick is to formulate carefully the {\bf induction hypothesis}. 
We suppose that we have constructed the chronological products 
$T_{p}(x_{1},\cdots,x_{p}), \quad p = 1, \dots, n - 1$
having the following properties: (\ref{sym}), (\ref{causality}) and
(\ref{unitarity}) for 
$p \leq n - 1$,
(\ref{causality}) and (\ref{commute}) for
$|X_{1}| + |X_{2}| \leq n - 1$
and
\be
gh(T_{p}) = 0
\label{gh-p}
\ee
for 
$p \leq n - 1$.

We also suppose that we have constructed the Wick polynomials
$T_{p/l}(x_{1},\cdots,x_{p}), \quad l = 1, \dots, p$
for
$p = 1, \dots, n - 1$
such that we have properties analogue to (\ref{sym}), (\ref{causality}) and
(\ref{unitarity}). We use a convention similar to (\ref{empty}): if 
$X = \{1,\dots,p\}$ we denote $T^{\mu}_{l}(X) \equiv
T^{\mu}_{p/l}(x_{1},\dots,x_{p}), \quad l \leq p$ 
and we assume that:
\be
T(\emptyset) \equiv {\bf 1}, \quad 
T^{\mu}_{l} (\emptyset) \equiv 0, \quad
T^{\mu}_{l} (X) \equiv 0, \quad {\rm for} \quad x_{l} \not\in X.
\label{empty-p/l}
\ee

Then the induction hypothesis is supplemented as follows.
\begin{itemize}

\item
Symmetry:
\be
T_{p/\pi(l)}(x_{\pi(1)},\cdots x_{\pi(p)}) = T_{p/l}(x_{1},\cdots x_{p}), 
\quad \forall \pi \in {\cal P}_{p}.
\label{sym-p/l}
\ee
for
$p = 1, \dots, n - 1$;

\item
Covariance with respect to
$SL(2,\C)$:
\be
U_{a, A} T^{\mu}_{p/l}(x_{1},\cdots, x_{p}) U^{-1}_{a, A} =
{\delta(A^{-1})^{\mu}}_{\rho}
T^{\rho}_{p}(\delta(A)\cdot x_{1}+a,\cdots, \delta(A)\cdot x_{p}+a), 
\label{invariance-p/l}
\ee
$p = 1, \dots, n - 1$;

\item
Charge conjugation:
\be
U_{C} T^{\mu}_{p/l}(x_{1},\cdots, x_{p}) U^{-1}_{C} =
T^{\mu}_{p/l}(x_{1},\cdots,x_{p}), 
\label{charge-p/l}
\ee
for
$p = 1, \dots, n - 1$;

\item
Causality
\be
T^{\mu}_{l}(X_{1}X_{2}) = 
T^{\mu}_{l}(X_{1}) T(X_{2}) + T(X_{1}) T^{\mu}_{l}(X_{2})
\quad \forall X_{1} \geq X_{2}
\label{causality-p/l}
\ee
and
\be
[T^{\mu_{1}}_{l_{1}}(X_{1}), T^{\mu_{2}}_{l_{2}}(X_{2})] = 0, \quad
[T^{\mu}_{l}(X_{1}), T(X_{2})] = 0
\quad {\rm if} \quad X_{1} \sim X_{2}
\label{commute-p/l}
\ee
for
$|X_{1}| + |X_{2}| \leq n - 1$.

\item
Unitarity; we introduce, in analogy to (\ref{antichrono}):
\bea
(-1)^{|X|} \bar{T}^{\mu}_{l}(X) \equiv \sum_{r=1}^{|X|} (-1)^{r} \sum
[ T^{\mu}_{l}(X_{1}) T(X_{2}) \cdots T(X_{r}) + 
\cdots 
\nonumber \\
+ T(X_{1}) \cdots T(X_{r-1})T^{\mu}_{l}(X_{r})] \quad
\label{antichrono-p/l}
\eea
where
$X_{1},\cdots,X_{r}$
is a partition of $X$ and we use in an essential way the convention
(\ref{empty-p/l}). 
We require
\be
\bar{T}^{\mu}_{l}(X) = T^{\mu}_{l}(X)^{\dagger}, \quad \forall X
\label{unitarity-p/l}
\ee
for
$|X| \leq n - 1$;

\item
Gauge invariance:
\be
d_{Q} T(X) = i \sum_{l} {\partial \over \partial x^{\mu}_{l}} 
T^{\mu}_{l}(X)
\label{gauge-n/p}
\ee
for all
$|X| \leq n - 1.$
The restriction 
$l \in X$
is not essential because of the convention (\ref{empty-p/l}).

\item
Ghost number:
\be
gh(T^{\mu}_{l}(X)) = 1
\label{gh-p/l}
\ee
for 
$|X| \leq n - 1$.
\end{itemize}

(ii) We observe that the induction hypothesis is valid for 
$p = 1$
according to (i). We suppose that it is true for 
$p \leq n - 1$
and prove it for
$p = n$.

First we establish in analogy to (\ref{unit}) that we have:
\be
\sum (-1)^{|X|} 
\left[ T^{\mu}_{l}(X) \bar{T}(Y) + T(X) \bar{T}^{\mu}_{l}(Y) \right] = 
0 = \sum (-1)^{|X|} 
\left[ \bar{T}^{\mu}_{l}(X) T(Y) + \bar{T}(X) T^{\mu}_{l}(Y) \right].
\label{unit-p/l}
\ee
where the sum goes over all partitions 
$X, Y$
with
$|X| + |Y| \leq n - 1$.

Also one has, similarly to (\ref{bar-causality}):
\be
\bar{T}^{\mu}_{l}(XY) = 
\bar{T}^{\mu}_{l}(Y) \bar{T}(X) + \bar{T}(Y) \bar{T}^{\mu}_{l}(X),
\quad \forall X \geq Y.
\label{bar-causality-p/l}
\ee
and
$|X| + |Y| \leq n - 1$.

Finally, from (\ref{gauge-n}) and the definitions of the antichronological
products
$T(X)$
and
$T^{\mu}(X)$
we have 
\be
d_{Q} \bar{T}(X) = i \sum_{l} {\partial \over \partial x^{\mu}_{l}} 
\bar{T}^{\mu}_{l}(X),
\label{bar-gauge-n/p}
\ee
for all
$|X| \leq n - 1$.

Now we can proceed in strict analogy with Subsection \ref{EG}. The proof of the
following items below goes in strict analogy to the proof of the similar
statements from the previous Subsection and can be easily provided with minimal
modifications. 
\begin{enumerate}

\item
One constructs from
$T(X), \quad T^{\mu}_{l}(X), \quad |X| \leq n - 1$
the expressions
$\bar{T}(X), \quad \bar{T}^{\mu}_{l}(X), \quad |X| \leq n - 1$
and proves the properties (\ref{bar-causality}) +
(\ref{bar-causality-p/l}) for
$|X| + |Y| \leq n - 1$
and (\ref{unit}) + (\ref{unit-p/l}) for
$ |X| \leq n - 1$.
\item
Beside lemma \ref{AR-prime} we have the following result:

\item
\begin{lemma}
Let us defines the expressions:
\be
A'^{\mu}_{n/l}(x_{1},\dots,x_{n-1};x_{n}) \equiv {\sum}' (-1)^{|Y|} 
\left[ T^{\mu}_{l}(X) \bar{T}(Y) + T(X) \bar{T}^{\mu}_{l}(Y) \right]
\ee
\be
R'^{\mu}_{n/l}(x_{1},\dots,x_{n-1};x_{n}) \equiv {\sum}' (-1)^{|Y|} 
\left[ \bar{T}^{\mu}_{l}(X) T(Y) + \bar{T}(X) T^{\mu}_{l}(Y) \right]
\ee
where the sum
${\sum}'$
goes over the partitions 
$
X \cup Y = \{1, \dots, n\}, \quad X \cap Y = \emptyset, \quad 
Y \not= \emptyset, \quad x_{n} \in X.
$

Now, let us suppose that we have a partition
$
P \cup Q = \{1, \dots, n - 1\}, \quad P \cap Q = \emptyset, 
\quad P \not= \emptyset.
$

Then:

If
$Qj \geq P$ one has:
\be
A'^{\mu}_{n/l}(x_{1},\dots,x_{n-1};x_{n}) = - 
\left[ T^{\mu}_{l}(Qn) T(P) + T(Qn) T^{\mu}_{l}(P) \right]
\ee
and if
$Qj \leq P$ one has:
\be
R'^{\mu}_{n/l}(x_{1},\dots,x_{n-1};x_{n}) = - 
\left[ T^{\mu}_{l}(P) T(Qn) + T(P) T^{\mu}_{l}(Qn) \right].
\ee
\label{AR-prime-n/l}
\end{lemma}

The proof is similar to the proof of lemma \ref{AR-prime} if one uses the
causality properties (\ref{causality-p/l}) and (\ref{bar-causality-p/l}).

\item
Beside corollary \ref{D} we have:
\begin{cor}
The expression
\be
D^{\mu}_{n}(x_{1},\dots,x_{n-1};x_{n}) \equiv 
A'_{n/l}(x_{1},\dots,x_{n-1};x_{n}) -
R'_{n/l}(x_{1},\dots,x_{n-1};x_{n}).
\label{com-D-n/l}
\ee
have causal support i.e.
$supp(D^{\mu}_{n}(x_{1},\dots,x_{n-1};x_{n})) 
\subset \Gamma^{+}(x_{n}) \cup \Gamma^{-}(x_{n})$.
\label{D-p/l}
\end{cor}

The proof goes exactly as the proof of the Corollary \ref{D}.

\item
In lemma \ref{wick} we must use the fact that
$dim(T_{1}) = 4$;
we also have the generalization:
\begin{lemma}
The distribution 
$D^{\mu}_{n}(x_{1},\dots,x_{n-1};x_{n})$
can be written as a sum
\be
D^{\mu}_{n}(x_{1},\dots,x_{n-1};x_{n}) = \sum_{i} 
d^{Q}_{i}(x_{1},\dots,x_{n-1};x_{n})
W^{Q}_{i}(x_{1},\dots,x_{n-1};x_{n})
\ee
with
$W^{Q}_{i}(x_{1},\dots,x_{n-1};x_{n})$
are linearly independent Wick monomials and 
$d^{Q}_{i}(x_{1},\dots,x_{n-1};x_{n})$
numerical distributions with causal support i.e
$
supp(d^{Q}_{i}(x_{1},\dots,x_{n-1};x_{n})) 
\subset \Gamma^{+}(x_{n}) \cup \Gamma^{-}(x_{n}).
$
Moreover, the set of Wick monomials appearing in the preceding formula can be
obtained from the expression
$
T_{1}(x_{1}) \cdots T^{\mu}_{l}(x_{i}) \cdots T_{n}(x_{n})
$
by taking all possible Wick contractions, eliminating the monomials for for
which the corresponding numerical distributions are factorizable and keeping a
linearly independent set.

Finally, the following limitations are valid:
\be
\omega(d^{Q}_{i}) + \omega(W^{Q}_{i}) \leq \omega(T_{1}) = 4, \quad \forall i.
\ee
\label{wick-p/l}
\end{lemma}

\item
\begin{cor}
There exists a 
$SL(2,\C)$-covariant causal splitting:
\be
D^{\mu}_{n/l}(x_{1},\dots,x_{n-1};x_{n}) =
A^{\mu}_{n/l}(x_{1},\dots,x_{n-1};x_{n}) - 
R^{\mu}_{n/l}(x_{1},\dots,x_{n-1};x_{n/l})
\label{decD-n/l}
\ee
with
$supp(A^{\mu}_{n/l}(x_{1},\dots,x_{n-1};x_{n})) \subset \Gamma^{+}(x_{n})$
and
$supp(R^{\mu}_{n/l}(x_{1},\dots,x_{n-1};x_{n})) \subset \Gamma^{-}(x_{n})$
for all
$l = 1, \dots, n.$
\end{cor}

The proof goes as in the case of the distribution
$D_{n}$
if one notices that the distributions
$d^{Q}_{i}(x_{1},\dots,x_{n-1};x_{n})$
defined above are 
$SL(2,\C)$-covariant. For this reason 
$A_{n} (R_{n})$
are called {\it advanced} (resp. {\it retarded}) products.

\item
Beside lemma \ref{unitD} we have 
\begin{lemma}
The following relation is true
\be
D^{\mu}_{n/l}(x_{1},\dots,x_{n-1};x_{n})^{\dagger} = (-1)^{n-1}
D^{\mu}_{n/l}(x_{1},\dots,x_{n-1};x_{n}).
\ee
In particular the causal splitting obtained above can be chosen such that
\be
A^{\mu}_{n/l}(x_{1},\dots,x_{n-1};x_{n})^{\dagger} = (-1)^{n-1}
A^{\mu}_{n/l}(x_{1},\dots,x_{n-1};x_{n}).
\ee
\label{unitD-p/l}
\end{lemma}
So, performing the substitutions:
\bea
A^{\mu}_{n/l}(x_{1},\dots,x_{n-1};x_{n}) \rightarrow {1\over 2}
\left[ A^{\mu}_{n/l}(x_{1},\dots,x_{n-1};x_{n})^{\dagger} + (-1)^{n-1}
A^{\mu}_{n/l}(x_{1},\dots,x_{n-1};x_{n})\right]
\nonumber \\
R^{\mu}_{n/l}(x_{1},\dots,x_{n-1};x_{n}) \rightarrow {1\over 2}
\left[ R^{\mu}_{n/l}(x_{1},\dots,x_{n-1};x_{n})^{\dagger} + (-1)^{n-1}
R^{\mu}_{n/l}(x_{1},\dots,x_{n-1};x_{n})\right] 
\eea
for all
$l = 1,\dots,n$
we do not affect the relation from the preceding corollary and we obtain a
causal splitting verifying the condition from the statement without spoiling 
the
$SL(2,\C)$-covariance.
$\qed$

\item
Now we have again theorem \ref{chronos} and also
\begin{thm}
Let us define
\bea
T^{\mu}_{n/l}(x_{1},\cdots, x_{n}) \equiv 
A^{\mu}_{n/l}(x_{1},\cdots, x_{n-1};x_{n}) - 
A'^{\mu}_{n/l}(x_{1},\cdots, x_{n-1};x_{n}) 
\nonumber \\ \equiv 
R^{\mu}_{n/l}(x_{1},\cdots, x_{n-1};x_{n}) - 
R'^{\mu}_{n/l}(x_{1},\cdots, x_{n-1};x_{n}).
\eea
Then these expressions satisfy the Poincar\'e covariance, causality and 
unitarity conditions (\ref{invariance-p/l}) (\ref{causality-p/l}) 
(\ref{commute-p/l}) and (\ref{unitarity-p/l}) for
$p = n$.
If we substitute
\bea
T_{n}(x_{1},\cdots, x_{n}) \rightarrow {1 \over n!}
\sum_{\pi} T_{n}(x_{\pi(1)},\cdots, x_{\pi(n)}),
\nonumber \\
T^{\mu}_{n/l}(x_{1},\cdots, x_{n}) \rightarrow {1 \over n!}
\sum_{\pi} T^{\mu}_{n/\pi^{-1}(l)}(x_{\pi(1)},\cdots, x_{\pi(n)})
\eea
where the sum runs over all permutations of the numbers
$\{1, \dots, n\}$
then we also have the symmetry axioms (\ref{sym}) and (\ref{sym-p/l}) for
$p = n.$
\label{chronos-n/l}
Finally, if we substitute
\be
T(X) \rightarrow {1 \over 2}\left[ T(X) + T(X)^{\dagger}\right],
\quad
T^{\mu}_{l}(X) \rightarrow {1 \over 2}
\left[ T^{\mu}_{l}(X) + (T(X)^{\mu}_{l})^{\dagger}\right]
\ee
we have charge conjugation invariance as well.
\end{thm}
\end{enumerate}

(iii) Now we investigate the possible obstruction to the extension of the
identity (\ref{gauge-n}) for
$|X| = n$.

\begin{prop}
The following relation is valid:
\be
d_{Q} T(X) = i \sum_{l} {\partial \over \partial x^{\mu}_{l}} 
T^{\mu}_{l}(X) + P(X), \quad |X| = n
\label{anoP}
\ee
where
$P(X) \equiv P_{n}(x_{1},\dots,x_{n})$
is a Wick polynomial (called anomaly) of the following structure:
\be
P_{n}(x) = \sum_{i} p_{i}(\partial) \delta^{n-1}(x) W^{Q}_{i}(x);
\label{wickP}
\ee
here 
$p_{i}$
are polynomials in the derivatives with the maximal degree restricted by
\be
deg(p_{i}) + \omega(W^{Q}_{i}) \leq 5.
\label{degP}
\ee

Moreover, we have the following properties:
\begin{enumerate}

\item
Symmetry
\be
P_{n}(x_{\pi(1)},\dots,x_{\pi(n)}) = P_{n}(x_{1},\dots,x_{n})
\label{symP}
\ee
for any permutation
$\pi \in {\cal P}_{n}$.

\item
$SL(2,\C)$-covariance:
\be
U_{a, A} P_{n}(x_{1},\cdots, x_{n}) U^{-1}_{a, A} =
P_{n}(\delta(A)\cdot x_{1}+a,\cdots, \delta(A)\cdot x_{n}+a), 
\quad \forall (a,A) \in inSL(2,\C).
\label{invariance-P}
\ee

\item
Charge conjugation invariance:
\be
U_{C} P_{n}(x_{1},\dots,x_{n}) U^{-1}_{C} =  P_{n}(x_{1},\dots,x_{n}).
\label{charge-P}
\ee

\item
Unitarity:
\be
P_{n}^{\dagger} \equiv (-1)^{n} P_{n}.
\label{uniP}
\ee

\item
Ghost numbers restrictions:
\be
gh_{+}(P_{n}) = 1, \quad gh_{-}(P_{n}) = 0.
\label{ghP}
\ee

\item
The polynomials
$p_{i}$
from the relation (\ref{wickP}) do not contain the matrix
$\gamma_{5}$.

\item
The Wick monomials
$W^{Q}_{i}$ 
from (\ref{wickP}) do not have field factors with coinciding points (i.e. of
the type
$
:\dots B_{1}(x)B_{2}(x) \dots :
$
for
$B_{i} = A_{\mu_{i}}, \quad i = 1,2$
or
$B_{1} = A_{\mu}, \quad B_{2} = u$.

\item
The Wick monomials
$W^{Q}_{i}$ 
from (\ref{wickP}) are linear in the fields
$u(x)$
or
$\partial_{\mu} u(x)$,
do not contain factors with derivatives of order
$\geq 2$
of the fields and at most one of the factors is differentiated.
\end{enumerate}
\label{anomaly-n}
\end{prop}
{\bf Proof:}
First we obtain from the lemmas \ref{AR-prime} and \ref{AR-prime-n/l} that:
\bea
d_{Q} A'_{n}(x_{1},\dots,x_{n-1};x_{n}) =  i \sum_{l=1}^{n}
{\partial \over \partial x^{\mu}_{l}} 
A'^{\mu}_{n/l}(x_{1},\dots,x_{n}),
\nonumber \\
d_{Q} R'_{n}(x_{1},\dots,x_{n-1};x_{n}) =  i \sum_{l=1}^{n}
{\partial \over \partial x^{\mu}_{l}} 
R'^{\mu}_{n/l}(x_{1},\dots,x_{n});
\eea
by substraction we get:
\be
d_{Q} D_{n}(x_{1},\dots,x_{n-1};x_{n}) =  i \sum_{l=1}^{n}
{\partial \over \partial x^{\mu}_{l}} 
D^{\mu}_{n/l}(x_{1},\dots,x_{n}),
\ee

We substitute here the causal decompositions (\ref{decD}) and (\ref{decD-n/l})
in the preceding relation and we get:
\bea
d_{Q} A_{n}(x_{1},\dots,x_{n-1};x_{n}) -  i \sum_{l=1}^{n}
{\partial \over \partial x^{\mu}_{l}} 
A^{\mu}_{n/l}(x_{1},\dots,x_{n}) =
\nonumber \\
d_{Q} R_{n}(x_{1},\dots,x_{n-1};x_{n}) -  i \sum_{l=1}^{n}
{\partial \over \partial x^{\mu}_{l}} 
R^{\mu}_{n/l}(x_{1},\dots,x_{n}).
\eea

Now we can reasons as in lemma \ref{cov-split} - see formula (\ref{cs}) : the 
left hand side has support in 
$\Gamma^{+}(x_{n})$
and the right hand side in
$\Gamma^{-}(x_{n})$
so the common value, denoted by
$P_{n}$
should have the support in
$\Gamma^{+}(x_{n}) \cap \Gamma^{-}(x_{n}) = \{x_{1} = \cdots = x_{n}\}$.
This means that we have:
\be
d_{Q} A_{n}(x_{1},\dots,x_{n-1};x_{n}) -  i \sum_{l=1}^{n}
{\partial \over \partial x^{\mu}_{l}} 
A^{\mu}_{n/l}(x_{1},\dots,x_{n-1};x_{n}) = P'_{n}(x_{1},\dots,x_{n-1};x_{n})
\ee
where
$P'_{n}$
has the structure (\ref{wickP}) from the statement. We now have immediately
the relation (\ref{anoP}) from the statement where
$P_{n}$
has the structure (\ref{wickP}). The limitation (\ref{degP}) follows rather
easily from the lemmas \ref{wick} and \ref{wick-p/l}. The same is true for the
ghost number restriction (\ref{ghP}). The restrictions (\ref{symP}) and 
(\ref{charge-P}) follow from the similar properties of the products
$T(X)$ 
and 
$T^{\mu}_{l}(X)$ from the preceding proposition. 

For unitarity we proceed as follows. First, we apply the BRST operator
$d_{Q}$
to the relation (\ref{unit}) with 
$Z \equiv XY $
of cardinal $n$.
If we use the induction hypothesis (\ref{gauge-n}) and (\ref{bar-gauge-n/p}) +
(\ref{anoP}) we get:
\be
d_{Q} \bar{T}(X) = i \sum_{l} {\partial \over \partial x^{\mu}_{l}} 
\bar{T}^{\mu}_{l}(X) + \bar{P}(X), \quad |X| = n
\label{bar-anoP}
\ee
where
\be
\bar{P}_{n} \equiv (-1)^{n-1} P_{n}.
\ee

Now we apply the conjugation $\dagger$ to the relation (\ref{gauge-n}), use the
relation (\ref{dag-w}) and compare to the relation (\ref{bar-anoP}): the
relation (\ref{uniP}) follows. 

The fact that the polynomials
$p_{i}$
from (\ref{wickP}) do not contain the matrix
$\gamma_{5}$
follows by induction: one proves in this way that all the numerical
distributions appearing in the expressions of
$D(X), \quad D^{\mu}_{l}(X), \quad |X| = n$
do not contain this matrix. 

The structure of the Wick monomials
$W^{Q}_{i}$ 
detailed in the last two items of the statement is proved by induction as well.
$\nabla$

(iv) There are a lot of restrictions on the anomaly
$P_{n}$
and we will be able to prove here that it can be chosen to be equal to $0$.
First, from the restrictions (\ref{degP})), the 
$SL(2,\C)$-covariance
(\ref{invariance-P}) and the restrictions from the last item of the preceding
proposition we obtain that
\be
P(X) = \sum_{i=1}^{17} {\cal P}_{i}(X)
\ee
where the list of the polynomials in the right hand side is:
\bea
{\cal P}_{1}(X) \equiv \sum_{ijl} 
:\bar{\psi}(x_{i}) p^{\rho}_{ijl}(X) \psi(x_{j}): \partial_{\rho} u(x_{l}),
\quad
{\cal P}_{2}(X) \equiv \sum_{ijl} :\left[ 
\partial_{\rho}\bar{\psi}(x_{i})\right] 
q^{\rho}_{ijl}(X) \psi(x_{j}): u(x_{l}),
\nonumber \\
{\cal P}_{3}(X) \equiv \sum_{ijl} :\bar{\psi}(x_{i}) r^{\rho}_{ijl}(X) 
\left[ \partial_{\rho} \psi(x_{j})\right]: u(x_{l}),
\quad
{\cal P}_{4}(X) \equiv \sum_{ijl} 
:\bar{\psi}(x_{i}) p_{ijl}(X) \psi(x_{j}): u(x_{l}),
\nonumber \\
{\cal P}_{5}(X) \equiv \sum_{ijkl} 
:\bar{\psi}(x_{i}) p^{\rho}_{ijkl}(X) \psi(x_{j}): A_{\rho}(x_{k}) u(x_{l}),
\quad
{\cal P}_{6}(X) \equiv \sum_{l} p_{l}(X) u(x_{l}),
\nonumber \\
{\cal P}_{7}(X) \equiv \sum_{kl} p^{\mu}_{kl}(X) \partial_{\mu} u(x_{l}),
\quad
{\cal P}_{8}(X) \equiv \sum_{kl} q^{\rho}_{kl}(X) A_{\rho}(x_{k}) u(x_{l}),
\nonumber \\
{\cal P}_{9}(X) \equiv \sum_{kl} p^{\rho\lambda}_{kl}(X) A_{\rho}(x_{k}) 
\partial_{\lambda} u(x_{l}),
\quad
{\cal P}_{10}(X) \equiv \sum_{kl} q^{\rho\lambda}_{kl}(X) 
\partial_{\lambda}A_{\rho}(x_{k})  u(x_{l}), \quad
\nonumber \\
{\cal P}_{11}(X) \equiv \sum_{klm} p^{\rho\lambda}_{klm}(X) 
:A_{\rho}(x_{k})A_{\rho}(x_{m}): u(x_{l}),
\nonumber \\
{\cal P}_{12}(X) \equiv \sum_{klm} p^{\rho\lambda\zeta}_{klm}(X) 
:A_{\rho}(x_{k})A_{\rho}(x_{m}): \partial_{\zeta}u(x_{l}),\quad
\nonumber \\
{\cal P}_{13}(X) \equiv \sum_{klm} p^{\rho\lambda\zeta}_{klm}(X) 
:A_{\rho}(x_{k}) (\partial_{\zeta}A_{\rho}(x_{m})): u(x_{l}),
\nonumber \\
{\cal P}_{14}(X) \equiv \sum_{klmr} p^{\rho\lambda\zeta}_{klmr}(X) 
:A_{\rho}(x_{k})A_{\rho}(x_{m})A_{\zeta}(x_{r}): u(x_{l}),
\nonumber \\
{\cal P}_{15}(X) \equiv \sum_{klmr} p^{\rho\lambda\zeta\tau}_{klmr}(X) 
:A_{\rho}(x_{k})A_{\rho}(x_{m})A_{\zeta}(x_{r}): \partial_{\tau}u(x_{l}),
\nonumber \\
{\cal P}_{16}(X) \equiv \sum_{klmr} q^{\rho\lambda\zeta\tau}_{klmr}(X) 
:A_{\rho}(x_{k})A_{\rho}(x_{m}) (\partial_{\tau}A_{\zeta}(x_{r})): u(x_{l}),
\nonumber \\
{\cal P}_{17}(X) \equiv \sum_{klmrs} p^{\rho\lambda\zeta\tau}_{klmrs}(X) 
:A_{\rho}(x_{k})A_{\rho}(x_{m})A_{\zeta}(x_{r})A_{\tau}(x_{s}): u(x_{l}),
\qquad
\label{anomalies}
\eea
where the expressions 
$p^{\dots}_{\dots}$
are numerical distributions which are 
$SL(2,\C)$-covariant and are also restricted by the following degree
conditions: 
\bea
deg(p^{\rho}_{ijl}), \quad  deg(q^{\rho}_{ijl}), \quad deg(r^{\rho}_{ijl}),
\quad deg(p^{\rho}_{ijkl}), \quad deg(p^{\rho\lambda\zeta\tau}_{klmr}) 
\quad deg(q^{\rho\lambda\zeta\tau}_{klmr}), \quad
deg(p^{\rho\lambda\zeta\tau}_{klmrs}) \leq 0, 
\nonumber \\
deg(p_{ijl}), \quad deg(p^{\rho\lambda\zeta}_{klm}), \quad
deg(q^{\rho\lambda\zeta}_{klm}), \quad deg(p^{\rho\lambda\zeta}_{klmr})  
\leq 1,
\nonumber \\
deg(p^{\rho\lambda}_{kl}),\quad deg(q^{\rho\lambda}_{kl}),
\quad deg(p^{\rho\lambda}_{klm}), 
\quad  \leq 2, 
\nonumber \\
deg(p^{\rho}_{kl}), \quad deg(q^{\rho}_{kl}) \leq 3,
\quad
deg(p_{l}) \leq 4. \qquad
\label{deg-p-i}
\eea
It is obvious that all these polynomials also verify individually all the
restrictions from the preceding proposition. In particular, charge conjugation
invariance gives immediately:
\be
{\cal P}_{i} = 0, \quad i = 6, 7, 11, 12, 13, 17.
\ee
We analyse now the other cases. The basic idea is to perform obvious
``integrations by parts" and exhibit the polynomials as follows:
\be
{\cal P}_{i}(X) =  i \sum_{l=1}^{n} {\partial \over \partial x^{\rho}_{l}}
N^{\rho}_{l}(X) + {\cal P}_{i}'(X)
\label{generic}
\ee
where the Wick monomials
$N^{\rho}_{l}$
verify the following equation:
\be
(N^{\rho}_{l})^{\dagger} =  (-1)^{n-1} N^{\rho}_{l}.
\label{hermite1}
\ee

Then we make the redefinition
\be
T^{\rho}_{n/l} \rightarrow T^{\rho}_{n/l} + N, \quad 
\bar{T}^{\rho}_{n/l} \rightarrow \bar{T}^{\rho}_{n/l} + (-1)^{n-1} N^{\rho}_{l}
\label{redef1}
\ee
without affecting the properties of the expressions
$T(X), \quad T^{\mu}_{l}(X), \quad |X| = n.$
In this way one can eliminate immediately some of the monomials from the list
(\ref{anomalies}). In this way one can eliminate:
\begin{itemize}
\item
${\cal P}_{7}$
by redefining
${\cal P}_{6}$;

\item
${\cal P}_{9}$
and
${\cal P}_{10}$
by redefining
${\cal P}_{8}$;

\item
${\cal P}_{13}$
by redefining
${\cal P}_{12}$;

\item
${\cal P}_{16}$
by redefining
${\cal P}_{15}$.
\end{itemize}
In all these cases the property (\ref{hermite1}) follows easily from
(\ref{uniP}).

We analyse now the remaining cases.

1) In this case we have the following generic form of the numerical
distribution: 
\be
p^{\rho}_{ijl}(X) = \sum_{ijl} K_{ijl} \gamma^{\rho} \delta^{n-1}(X).
\label{1}
\ee
This means that we have:
\be
{\cal P}_{1}(X) = K \delta^{n-1}(X) \quad 
:\bar{\psi}(x_{n}) \gamma^{\rho} \psi(x_{n}):  \partial_{\rho} u(x_{n}).
\ee
From (\ref{uniP}) one finds out that
\be
\bar{K} = (-1)^{n} K.
\label{K}
\ee
Now one sees that
\be
{\cal P}_{1} = d_{Q} N
\label{coboundary}
\ee
where
\be
N(X) = K \delta^{n-1}(X) \quad 
:\bar{\psi}(x_{n}) \gamma^{\rho} \psi(x_{n}):  A_{\rho}(x_{n})
\ee
verifies, because of (\ref{K}), the relation
\be
N(X)^{\dagger} = (-1)^{n-1} N(X).
\label{hermite2}
\ee
It means that we can make the substitutions
\be
T(X) \rightarrow T(X) + N(X), \quad 
\bar{T}(X) \rightarrow \bar{T}(X) + (-1)^{n-1} N(X)
\label{redef2}
\ee
without modifying the properties of the chronological products. But in this way
we will have
${\cal P}_{1} = 0.$

2), 3) 4) In the first two cases the structure of the numerical distributions
$q$ and $r$ is similar to the structure (\ref{1}) above. If we use in the first
two cases Dirac equation (\ref{dirac-equ}) it follows that the
sum of these two contributions is of the form:
\be
{\cal P}_{0}(X) = K \delta^{n-1}(X) \quad 
:\bar{\psi}(x_{n}) \psi(x_{n}): u(x_{n}).
\label{0}
\ee
In the case 4) we have the generic form
\be
p_{ijl}(X) = \sum_{ijl} \left( K_{ijl} {\bf 1} 
+ K_{ijlm} \gamma\cdot\partial_{m}\right) \delta^{n-1}(X).
\label{4}
\ee
(where in the sum the values of the indices $i,j,k$ must be distinct..) We
integrate by parts, make the redefinition (\ref{redef1}) and conclude that we
can take
\be
{\cal P}_{4}(X) = \delta^{n-1}(X) W(x_{n})
\label{generic2}
\ee
where the Wick monomial $W$ is of the following type:
\be
W(x) = \{ K_{1} :[\partial_{\rho}\bar{\psi}(x)] \gamma^{\rho} \psi(x): 
+ K_{2} :\bar{\psi}(x) \gamma^{\rho} [\partial_{\rho}\psi(x)]:  \}u(x)
+ K_{3} :\bar{\psi}(x) \gamma^{\rho} \psi(x): \partial_{\rho}u(x)
\ee
In the first two terms we use Dirac equation (\ref{dirac-equ}). In the end we
obtain that
${\cal P}_{4}$
can be taken of the form
${\cal P}_{0}$.
Now it follows by elementary computations that charge
conjugation invariance imposes
${\cal P}_{0} = 0$.

5) In this case, the numerical distributions
$p^{\dots}_{\dots}$
have the same structure as in the case 1) so we end up with
\be
{\cal P}_{5}(X) = \delta^{n-1}(X) \quad 
K :\bar{\psi}(x_{n}) \gamma^{\rho} \psi(x_{n}):  A_{\rho}(x_{n}) u(x_{n}).
\ee
This contribution is zero because of the charge conjugation invariance.

8) In this case we have
\be
p^{\mu}(X)  = \sum_{i} \left( C_{i} \partial_{i}^{\mu} 
+ \sum_{ijk} D_{ijk} \partial_{i}^{\mu} \partial_{j}\cdot\partial_{k}
+ \varepsilon^{\mu\nu\rho\lambda} D'_{ijk} 
\partial_{i\nu} \partial_{j\rho} \partial_{k\lambda} \right)
\delta^{n-1}(X).
\label{8}
\ee
By  integrations by parts it follows that we can take
${\cal P}_{8} = 0$
of the form (\ref{generic2}):
\be
{\cal P}_{8}(X) = \delta^{n-1}(X) W(x_{n})
\ee
where the Wick monomial $W$ is of the following type (use of the equations of
motion must be made):
\be
W(x) = C_{1} (\partial^{\mu}A_{\mu}) u 
+ C_{2} A_{\mu} \partial^{\mu}u 
+ C_{3} (\partial^{\mu}A^{\mu}) (\partial_{\mu}\partial_{\nu}u)
+ C_{4} (\partial^{\mu}\partial^{\nu} A_{\mu})(\partial_{\nu}u)
\ee
which can be exhibited in the form (\ref{coboundary}):
\be
W = d_{Q} N
\ee
with
\be
N = i \left[ C_{1} :\tilde{u} u: - {1\over 2} C_{2} :A_{\mu} A^{\mu}:
- {1\over 2} C_{3} :(\partial_{\nu}A_{\mu})(\partial^{\nu} A^{\mu}):
+ C_{4} :(\partial_{\mu}\tilde{u})\partial^{\mu} u):\right]
\ee
So, we can make the redefinition (\ref{redef2}) and put
${\cal P}_{8} = 0$.

14) In this case we have
\be
p^{\mu\nu\rho}(X) =  \sum_{i} \left( C_{i} g^{\mu\nu} \partial_{i}^{\rho}
+ D_{i} g^{\mu\rho} \partial_{i}^{\nu}
+ E_{i} g^{\rho\nu} \partial_{i}^{\mu}
+ \varepsilon^{\mu\nu\rho\lambda} F_{i} 
\partial_{i\lambda}\right) \delta^{n-1}(X).
\label{14}
\ee
If we integrate by parts, we get a formula of the type (\ref{generic2})
\be
{\cal P}_{14}(X) = \delta^{n-1}(X) W(x_{n})
\ee
with $W$ of the following form:
\be
W(x) =  K_{1} :A_{\rho}A^{\rho}(\partial^{\mu}A_{\mu}): u
+ K_{2} :A_{\rho}A^{\rho}A_{\mu}: \partial^{\mu}u
+ K_{3} :A_{\rho}A_{\mu}(\partial^{\mu}A^{\rho}): u
\ee
On can get rid of the last term by redefining the first two ones. But in the
case 
$K_{3} = 0$
we have
\be
W = d_{Q} N
\ee
with
\be
N = i \left( K_{1} :A^{\mu}A_{\mu} \tilde{u} u: 
- {1\over 4} K_{2} :A_{\mu} A^{\mu}A_{\rho} A^{\rho}:\right)
\ee
so, we can make 
${\cal P}_{14} = 0$
as in the case 8).

15) In this case we have
\be
p^{\mu\nu\rho\lambda}(X) =  \left( K_{1} g^{\mu\nu} g^{\rho\lambda}
+ K_{2} g^{\mu\rho} g^{\nu\lambda} + K_{3} g^{\mu\lambda} g^{\rho\nu}
+ K_{4} \varepsilon^{\mu\nu\rho\lambda} \right) \delta^{n-1}(X).
\label{15}
\ee
so
\be
{\cal P}_{12}(X) =  C \delta^{n-1}(X) W(x_{n})
\ee
where $W$ is of the form (\ref{coboundary}) with
\be
N = - {i \over 4} :A_{\mu} A^{\mu} A_{\rho} A^{\rho}:
\ee
and we can proceed as at 1).

In conclusion, we can make in (\ref{gauge-n})
$P_{n} = 0$.
This finishes the proof.
$\qed$
\begin{rem}
We can fix the covariance properties with respect to the spatial and temporal
inversions as in \cite{Sc1} ch. 4.4.
\end{rem}

We now determine the non-unicity of the chronological products
$T(X)$. We have:
\begin{prop}
Suppose that
$T(X)$ and $T'(X)$
are two solutions of the renormalization theory for quantum electrodynamics,
verifying gauge invariance in the sense of the preceding theorem and the power
counting condition (\ref{deg-t}). Then we have
\be
T(X) - T'(X) = d_{Q} N(X) 
+ i \sum_{l \in X} {\partial \over \partial x^{\mu}_{l}} N^{\mu}_{l}(X) 
+ C \delta^{n-1}(x) :\bar{\psi}(x_{n}) \gamma^{\rho} \psi(x_{n}):
A_{\mu}(x_{n}) 
\label{ren}
\ee
with 
$i^{n} C \in \R$. 
In particular, we can absorb the last term in the interaction Lagrangian by
redefining the coupling constant up to order $n$.
\end{prop}

{\bf Proof:}
From the gauge invariance condition, the expression
$F(X) \equiv T(X) - T'(X)$
verifies:
\be
d_{Q} F(X) = i \sum_{l \in X} {\partial \over \partial x^{\mu}_{l}} 
F^{\mu}_{l}(X)
\ee
for some Wick polynomials
$F^{\mu}_{l}(X)$.
Now we have from lemma \ref{wick}
\be
F(X) = \sum_{i} f_{i}(x) W_{i}(x)
\ee
with the numerical distributions of the form
\be
f_{i}(x) = p_{i}(\partial) \delta^{n-1}(x)
\ee
where 
$p_{i}$
are polynomials verifying the restrictions
\be
deg(p_{i}) + \omega(W_{i}) \leq \omega(T_{1}) = 4, \quad \forall i.
\label{deg-f}
\ee

We also have all the properties of symmetry, covariance with respect to 
$SL(2,\C)$
and charge conjugation invariance.  We list all polynomials fulfilling these
requirements and we obtain the result.  $\qed$
\begin{rem}
The possibility of redefining the charge such that the last term in (\ref{ren})
disappears is named charge renormalization. We see that, in a rigorous version
of the renormalization theory there is no mass renormalization, because one
starts from the very beginning in a definite Fock space where the Dirac particle
has a fixed mass m.
\end{rem}
\begin{rem}
It is plausible that some descent equations of the type:
\be
d_{Q} T^{\mu_{1},\dots,\mu_{k}}_{l_{1},\dots,l_{k}} =
i \sum_{l_{k+1} \not= l_{1},\dots,l_{k}} 
{\partial \over \partial x^{\mu_{k+1}}_{k+1}} 
T^{\mu_{1},\dots,\mu_{k+1}}_{l_{1},\dots,l_{k+1}}, \quad k \leq n
\ee
can be established by refining the induction hypothesis. Such formul\ae~ could
be useful in more complicated theories.
\end{rem}

We close this Section with a comparison between our proof and the proof
appearing in \cite{Sc1}, ch. 4.6. In this reference one works in a quantization
formalism for the electromagnetic field without ghosts. One can prove, also by
induction,  a more precise formula for the chronological products:
\be
T_{n}(x_{1},\dots,x_{n}) \sum_{I,J,K} 
:\prod_{i \in I} \bar{\psi}(x_{i}) t^{\mu_{K}}_{I,J,K}(X)
\prod_{j \in J} \psi(x_{j}): :\prod_{k \in K} A_{\mu_{k}}(x_{k}):
\ee
where: a) the sum runs over all distinct triplets 
$I, J, K \subset \{1,\dots,n\}$
verifying
$|I| = |J|$;
b) by 
$\mu_{K}$
we mean the set
$\{\mu_{k}\}_{k \in K}$;
c) the expression
$t^{\mu_{K}}_{I,J,K}$
are numerical distributions (in fact, they take values in the matrix space
$M_{\C}(4,4)^{\otimes |I|}$.)

The possibility of having non-zero intersection between the sets $I$, $J$ and
$K$ is permitted. The condition of gauge invariance (\ref{gauge-n}) can be
translated into conditions on these numerical distributions
$t^{\mu_{K}}_{I,J,K}$
which are the famous Ward-Takahashi identities. They have a rather complicated
form precisely because of the possible non-void intersections mentioned above.
In fact, the proof from \cite{Sc1} relies on the following identities for which
we did not found an elementary proof. Formula (4.6.36) from this reference is 
a particular case of the formul\ae~ below.
\bea
t^{\rho\mu_{K}}_{iI,J,iK}(X) = e \sum_{j \not\in I} \left[
\gamma^{\rho} S^{F}_{m}(x_{i}-x_{j}) \otimes \cdots \otimes {\bf 1}\right]
t^{\mu_{K}}_{jI,J,K}(X \ x_{i}), \quad \forall i \not\in I, |J| < n,
\nonumber \\
t^{\mu_{K}\rho}_{I,Jj,Kj}(X) = e \sum_{i \not\in J} 
t^{\mu_{K}}_{jI,JK,}(X \ x_{j}) \left[{\bf 1} \otimes \cdots \otimes
S^{F}_{m}(x_{i}-x_{j}) \gamma^{\rho}\right], 
\quad \forall j \not\in J, |I| < n,
\nonumber \\
t^{\mu_{K}}_{iI,iJ,K}(X) = e \sum_{l \not\in K} 
D^{F}_{0}(x_{i}-x_{l}) t^{\mu_{K}\rho}_{I,J,Kl}(X \ x_{l}) \otimes
\gamma_{\rho} \quad \forall i \not\in I \cup J, |I| < n.
\eea

\section{Renormalizability of the Scalar Quantum Electrodynamics}{\label{sqed}}

\subsection{The Interaction Lagrangian}{\label{int-sqed}}

In this case, the matter field is a complex scalar field 
$\phi$.
The Hilbert space of the model is generated by applying on the vacuum 
$\Phi_{0}$ 
the free fields 
$
A^{\mu}(x), \quad u(x), \quad \tilde{u}(x), \quad \phi(x)
$
and
$\overline{\phi}(x)$.
We completely characterize these fields as in Subsection \ref{int-qed}. The
electromagnetic field and the corresponding ghost fields are determined by the
same relations as in this Subsection. 

\begin{itemize}

\item
Equation of motion; beside (\ref{equ}) and (\ref{dirac-equ}) we have
Klein-Gordon equations for the fields $\phi$ and $\bar{\phi}$:
\be
(\square + m^{2}) \phi(x) = 0,\quad (\square + m^{2}) \bar{\phi}(x) = 0.
\label{KG}
\ee

\item
Canonical (anti)commutation relations; beside (\ref{CCR}) we require:
\bea
[A^{\mu}(x),\phi(y)] = 0, \quad
[\phi(x),u(y)] = 0, \quad
[\phi(x),\tilde{u}(y)] = 0
\nonumber \\
~[\phi(x), \phi(y)] = 0, \quad
[\overline{\phi}(x), \overline{\phi}(y)] = 0, \quad
[\phi(x), \overline{\phi}(y)] = D_{m}(x-y) \times {\bf 1}.
\label{CCR-m}
\eea

\item
$SL(2,\C)$-covariance. We keep the corresponding relations (\ref{poincare}) 
from Subsection \ref{zero} and we add:
\be
U_{a,A} \phi(x) U^{-1}_{a,A} = \phi(\delta(A) \cdot x + a).
\ee

\item
Spatio-temporal invariance; beside (\ref{spatial}) and (\ref{temporal}) we
impose: 

\be
U_{I_{s}} \phi(x) U_{I_{s}}^{-1} = \phi(I_{s}\cdot x),
\quad
U_{I_{t}} \phi(x) U_{I_{t}}^{-1} = \overline{\phi}(I_{t}\cdot x).
\label{PT}
\ee

\item
Charge invariance. The unitary operator realizing the charge conjugation
verifies (\ref{charge}) and
\be
U_{C} \phi(x) U_{C}^{-1} = \overline{\phi}(x).
\label{charge-m}
\ee

\item
Moreover, we suppose that all these operators are leaving the vacuum invariant
i.e. we have (\ref{inv-vacuum}).
\end{itemize}

As in the case of spinorial electrodynamics, one can show that the operators
$
U_{a,A}, \quad U_{I_{s}}, \quad U_{I_{t}}
$
and
$U_{I_{st}}$
are a representation of the Poincar\'e group; however, in this case the
operators realizing the discrete symmetries square to identity. Moreover, the
charge conjugation operator commutes with all the preceding operators.

We give as before in
${\cal H}^{gh}$
the sesqui-linear form
$<\cdot,\cdot>$
which is completely characterize by requiring beside (\ref{conjugate}):
\be
\phi(x)^{\dagger} = \overline{\phi}(x).
\ee

The expression of the supercharge gets remains the same (\ref{supercharge})
and one can see that (\ref{Q-0}) stays true; to (\ref{Q-com}) one must add:
\be
[Q, \phi] = 0, \quad [Q, \overline{\phi}] = 0.
\label{Q-com-m}
\ee

Alternatively if ${\cal W}$ is the linear space of all Wick monomials acting 
in the Fock space
${\cal H}^{gh}$
containing the fields
$
A_{\mu}(x),\quad u(x),\quad \tilde{u}(x), \quad \phi(x)
$
and
$\overline{\phi}(x)$
then the expression of the BRST operator is determined by
\bea
d_{Q} u = 0, \quad d_{Q} \tilde{u} = - i \partial^{\mu} A_{\mu}, 
\quad d_{Q} A_{\mu} = i \partial_{\mu} u, \quad
d_{Q} \phi = 0, \quad d_{Q} \overline{\phi} = 0.
\label{BRSTm}
\eea

The relations (\ref{uw}) and (\ref{UQ}) valid in this case also.

As in the case of spinorial QED we give the expression of the interaction
Lagrangian. 
\be
T_{1}(x) \equiv
i e :\overline{\phi}(x) \stackrel{\leftrightarrow}{\partial_{\mu}} \phi(x): 
A^{\mu}(x).
\label{T1-s}
\ee
(here $e$ is a real constant called the electron charge) and can verify easily
that the properties (\ref{inv1}), (\ref{causality1}) and (\ref{unitarity1}) are
true. We also have instead of (\ref{pct-qed}):
\be
U_{I_{s}} T_{1}(x) U^{-1}_{I_{s}} = - T_{1}(I_{s}\cdot x),\quad
U_{I_{t}} T_{1}(x) U^{-1}_{I_{t}} = T_{1}(I_{t}\cdot x),\quad
U_{C} T_{1}(x) U^{-1}_{C} = T_{1}(x).
\label{pct-qed-s}
\ee

As in the case of spinorial electrodynamics we have first order gauge
invariance i.e. (\ref{gauge2}) for 
$n = 1$ 
with
\be
T_{1/1}^{\mu}(x) \equiv 
i e :\bar{\phi}(x) \stackrel{\leftrightarrow}{\partial^{\mu}} \phi(x): u(x)
\label{T1/1-s}
\ee
and we have, as in the case of spinorial QED, the relations (\ref{gh1/1}),
(\ref{causality1/1}) and (\ref{unitarity1/1}); instead of (\ref{inv1/1}) we
have: 
\bea
U_{a,A} T^{\mu}_{1/1}(x) U^{-1}_{a,A} = {\delta(A^{-1})^{\mu}}_{\rho}
T^{\rho}_{1/1}(\delta(A)\cdot x+a), \quad \forall A \in SL(2,\C),
\nonumber \\
U_{I_{s}} T^{\mu}_{1/1}(x) U^{-1}_{I_{s}} = - {(I_{s})^{\mu}}_{\rho}
T^{\rho}_{1/1}(I_{s}\cdot x),\quad
U_{I_{t}} T^{\mu}_{1/1}(x) U^{-1}_{I_{t}} = {(I_{s})^{\mu}}_{\rho}
T^{\rho}_{1/1}(I_{t}\cdot x), \quad
\nonumber \\
U_{C} T^{\mu}_{1/1}(x) U^{-1}_{C} = T^{\mu}_{1/1}(x).
\label{inv1/1-s}
\eea

We note the change of sign in the relations describing the behaviour of
spinorial QED  and scalar QED with respect to spatial inversion.

\subsection{Gauge Invariance of the Scalar QED}

We prove that scalar QED as defined in the previous Subsection is
gauge invariant, i.e. the $S$-matrix is factorizable (in the adiabatic limit)
to the physical Hilbert space; the proof will be extremely similar to the proof
from Subsection \ref{ren-qed} and we will indicate only the appropriate
changes.

\begin{thm}
Suppose that
$T_{1}$
is given by (\ref{T1-s}) above. Then the distributions
$T_{n}$
can be constructed such that, beside the conditions of symmetry,
$SL(2,\C)$-covariance,
causality and unitarity (\ref{sym}), (\ref{invariance}), (\ref{causality}), 
(\ref{unitarity}), verify the condition of P and T covariance, charge
invariance and gauge invariance: 
\be
U_{I_{s}} T_{n}(x_{1},\dots,x_{n}) U^{-1}_{I_{s}} = (-1)^{n}
T_{n}(I_{s}\cdot x_{1},\dots,I_{s}\cdot x_{n}),
\label{spatial-n-s}
\ee
\be
U_{I_{t}} T_{n}(x_{1},\dots,x_{n}) U^{-1}_{I_{t}} = 
\bar{T}_{n}(I_{t}\cdot x_{1},\dots,I_{t}\cdot x_{n}),
\label{temporal-n-s}
\ee
\be
U_{C} T_{n}(x_{1},\dots,x_{n}) U^{-1}_{C} =  T_{n}(x_{1},\dots,x_{n}),
\label{charge-n-s}
\ee
\be
d_{Q} T_{n}(x_{1},\dots,x_{n}) =  i \sum_{l=1}^{n}
{\partial \over \partial x^{\mu}_{l}} 
T^{\mu}_{n/l}(x_{1},\dots,x_{n}),
\quad \forall n \in \N^{*}
\label{gauge-n-s}
\ee
where
$T^{\mu}_{n/l}, \quad l = 1,\dots,n$
are some Wick monomials.
\label{gauge-QED-s}
\end{thm}
{\bf Proof:}

(i) The induction hypothesis is essentially the same as in the case of
spinorial QED. We note the sign difference in the relation expression spatial
inversion covariance. A similar difference appears in the induction hypothesis
for the products
$T^{\mu}_{l}(X)$.

We observe that the induction hypothesis is valid for 
$p = 1$
according to the results from the previous Subsection. We suppose that it is 
true for 
$p \leq n - 1$
and prove it for
$p = n$.
We can proceed in strict analogy with the proof from Subsection \ref{ren-qed}.
Everything stays unchanged with minor modification. The anomaly
$P_{n}$
is constructed in the same way and is constrained by the following conditions.
\begin{itemize}

\item
It has the polynomial structure 
\be
P_{n}(x) = \sum_{i} p_{i}(\partial) \delta^{n-1}(x) W^{Q}_{i}(x);
\label{wickP-ym}
\ee
here 
$p_{i}$
are polynomials in the derivatives and
$W_{i}$
are Wick monomials in all the free fields of the theory. We have the degree
restriction
\be
deg(p_{i}) + \omega(W^{Q}_{i}) \leq 5,
\label{degP-ym}
\ee
where we compute the degree of a Wick monomial by attributing to every field or
derivative the value $1$.

\item
$SL(2,\C)$-covariance: it is unchanged
\be
U_{a, A} P_{n}(x_{1},\cdots, x_{n}) U^{-1}_{a, A} =
T_{n}(\delta(A)\cdot x_{1}+a,\cdots, \delta(A)\cdot x_{n}+a), 
\quad \forall (a,A) \in inSL(2,\C).
\label{invariance-P-s}
\ee

\item
space-time covariance:
\be
U_{I_{s}} P_{n}(x_{1},\dots,x_{n}) U^{-1}_{I_{s}} = 
(-1)^{n} P_{n}(I_{s} \cdot x_{1},\dots, I_{s} \cdot x_{n}),
\label{spatial-s}
\ee
\be
U_{I_{t}} P_{n}(x_{1},\dots,x_{n}) U^{-1}_{I_{t}} = 
(-1)^{n-1} P_{n}(I_{t} \cdot x_{1},\dots, I_{t} \cdot x_{n}).
\label{temporal-s}
\ee

\item
charge conjugation invariance:
\be
U_{C} P_{n}(x_{1},\dots,x_{n}) U^{-1}_{C} = P_{n}(x_{1},\dots,\cdot x_{n}).
\label{charge-s}
\ee

\item
Unitarity:
\be
P_{n}^{\dagger} \equiv (-1)^{n} P_{n}.
\label{uniP-s}
\ee

\item
Ghost numbers restrictions:
\be
gh(P_{n}) = 1.
\label{ghP-s}
\ee
\item
The last two items of Proposition \ref{anomaly-n} are essentially the same (the
only modification is that we can two derivatives on the scalar fields).
\end{itemize}
\label{anomaly-n-s}

(ii) The list of possible anomalies consists of two types of terms: (a) the
expressions 
${\cal P}_{6} - {\cal P}_{17}$
from (\ref{anomalies}); (b) anomalies containing a pair
$:\overline{\phi} \phi:$
with possible derivatives. We list these terms below:
\bea
{\cal P}'_{1}(X) \equiv \sum_{ijl} p^{(1)\mu\nu}_{ijl}(X)
:\partial_{\mu}\bar{\phi}(x_{i})  \phi(x_{j}): \partial_{\nu} u(x_{l}),
\quad
{\cal P}'_{2}(X) \equiv \sum_{ijl} p^{(2)\mu\nu}_{ijl}(X)
:\phi(x_{i}) \partial_{\mu}\phi(x_{j}): \partial_{\nu} u(x_{l}),
\nonumber \\
{\cal P}'_{3}(X) \equiv \sum_{ijl}  p^{(3)\mu\nu}_{ijl}(X) 
:\partial_{\mu}\bar{\phi}(x_{i}) \partial_{\nu}\phi(x_{j}): u(x_{l}),
\quad
{\cal P}'_{4}(X) \equiv \sum_{ijl} p^{(4)\mu\nu}_{ijl}(X)
:\partial_{\mu}\partial_{\nu}\bar{\phi}(x_{i})  \phi(x_{j}): u(x_{l}),
\nonumber \\
{\cal P}'_{5}(X) \equiv \sum_{ijl} p^{(5)\mu\nu}_{ijl}(X)
:\bar{\phi}(x_{i})  \partial_{\mu}\partial_{\nu}\phi(x_{j}): u(x_{l}),
\quad
{\cal P}'_{6}(X) \equiv \sum_{ijl} p^{(6)\mu}_{ijl}(X)
:\bar{\phi}(x_{i})  \phi(x_{j}): \partial_{\mu}u(x_{l}),
\nonumber \\
{\cal P}'_{7}(X) \equiv \sum_{ijl} p^{(7)\mu}_{ijl}(X)
:\bar{\phi}(x_{i}) \partial_{\mu}\phi(x_{j}): u(x_{l}),
\quad
{\cal P}'_{8}(X) \equiv \sum_{ijl}  p^{(8)\mu}_{ijl}(X) 
:\bar{\phi}(x_{i}) \partial_{\mu}\phi(x_{j}): u(x_{l}),
\nonumber \\
{\cal P}'_{9}(X) \equiv \sum_{ijl} p^{(9)}_{ijl}(X)
:\bar{\phi}(x_{i})  \phi(x_{j}): u(x_{l}),
\nonumber \\
{\cal P}'_{10}(X) \equiv \sum_{ijkl} p^{(10)\mu\nu}_{ijkl}(X)
:\bar{\phi}(x_{i}) \phi(x_{j}): A_{\mu}(x_{k}) \partial_{\nu}u(x_{l}),
\nonumber \\
{\cal P}'_{11}(X) \equiv \sum_{ijkl} p^{(11)\mu\nu}_{ijkl}(X)
:\bar{\phi}(x_{i}) \phi(x_{j}): \partial_{\nu}A_{\mu}(x_{k}) u(x_{l}),
\nonumber \\
{\cal P}'_{12}(X) \equiv \sum_{ijkl} p^{(12)\mu\nu}_{ijkl}(X)
:\partial_{\nu}\bar{\phi}(x_{i}) \phi(x_{j}): A_{\mu}(x_{k}) u(x_{l}),
\nonumber \\
{\cal P}'_{13}(X) \equiv \sum_{ijkl} p^{(13)\mu\nu}_{ijkl}(X)
:\bar{\phi}(x_{i}) \partial_{\nu}\phi(x_{j}): A_{\mu}(x_{k}) u(x_{l}),
\nonumber \\
{\cal P}'_{14}(X) \equiv \sum_{ijkl} p^{(10)\mu}_{ijkl}(X)
:\bar{\phi}(x_{i}) \phi(x_{j}): A_{\mu}(x_{k}) u(x_{l}). \qquad
\label{anomalies-s}
\eea
Degree limitations of the type (\ref{deg-p-i}) are still valid and are easy to
write down. The anomalies of the type (a) are treated exactly as in the
Subsection \ref{ren-qed}. For the rest, we outline the arguments, which are
quite similar to the previous ones.

First, we use integrations by parts to obtain formul\ae~ of the type
(\ref{generic}) and make the redefinition (\ref{redef1}). In this way we can
get rid of the following terms:
\begin{itemize}
\item
${\cal P}'_{1}$
and
${\cal P}'_{2}$
by redefining
${\cal P}'_{3} - {\cal P}'_{5}$;

\item
${\cal P}'_{6}$
by redefining
${\cal P}'_{7}$
and
${\cal P}'_{8}$;

\item
${\cal P}_{11}$
by redefining
${\cal P}'_{10}, \quad {\cal P}'_{13}, \quad {\cal P}'_{14}$.
\end{itemize}

We give the analysis of the remaining cases.

3') In this case the generic form of the distribution: 
$p^{\mu\nu}_{ijl}$
is 
\be
p^{\mu\nu}_{ijl}(X) = g^{\mu\nu} p_{ijl} \delta^{n-1}(X).
\label{p1}
\ee
This means that we have:
\be
{\cal P}'_{3} = \delta^{n-1}(X) \quad 
:\partial^{\mu}\bar{\phi}(x_{n}) \partial_{\mu} \phi(x_{n}): u(x_{n}).
\ee
But one proves rather elementary that from charge conjugation invariance 
we have
$
{\cal P}'_{3} = 0.
$

4'), 5') The structure of the numerical distribution is again (\ref{p1}); we
can use in this case Klein-Gordon equation (\ref{KG}) and we obtain expressions
of the type 9'). If we redefine 
${\cal P}'_{9}$
we eliminate this term.

7')-8') In these cases the structure of the numerical distribution is
\be
p^{\mu}_{ijl}(X) = \sum_{m} p_{ijlm} \partial^{\mu}_{m} \delta^{n-1}(X).
\label{p2}
\ee
By integrations by parts, we get two types of terms: contributions of the type
3') which disappear because charge conjugation invariance and contributions of
the type 9'). If we redefine 
${\cal P}'_{9}$
we eliminate these terms.

9') The generic form of the numerical distribution is
\be
p_{ijl}(X) = c_{0} \delta^{n-1}(X) + 
\sum_{m} p_{ijlm} \square_{m} \delta^{n-1}(X).
\label{p3}
\ee

After integration by parts and utilization of the equations of motion we get an
expression of the type (\ref{generic}); after making the redefinition
(\ref{redef1}) we remain with an expression of the type:
\be
{\cal P}'_{9} = \delta^{n-1}(X) \quad \left[
c_{1} :\bar{\phi}(x_{n}) \phi(x_{n}): +
c_{2} :\partial^{\mu}\bar{\phi}(x_{n}) \partial_{\mu} \phi(x_{n}): \right]
u(x_{n});
\ee
this expression is zero because of charge conjugation invariance.

10'), 12'), 13') In these cases the numerical distributions are of the generic
form (\ref{p1}) so we have the following expressions:
\bea
{\cal P}'_{10} = c_{1} \delta^{n-1}(X) :\bar{\phi}(x_{n}) \phi(x_{n}): 
A_{\mu}(x_{n}) \partial^{\mu}u(x_{n}),
\nonumber \\
{\cal P}'_{12} = c_{2} \delta^{n-1}(X) 
:\partial^{\mu}\bar{\phi}(x_{n}) \phi(x_{n}): A_{\mu}(x_{n}) u(x_{n}),
\nonumber \\
{\cal P}'_{13} = c_{3} \delta^{n-1}(X) 
:\bar{\phi}(x_{n}) \partial^{\mu}\phi(x_{n}): A_{\mu}(x_{n}) u(x_{n}),
\eea

From the unitarity requirement (\ref{uniP-s}) we obtain
$c_{2} = c_{3}$.
This observation allows us group the last two contributions and make some
integration by parts; as a result we obtain a contribution of the type 
${\cal P}'_{10}$ 
and one of the type
\be
{\cal P}'_{11} = c_{2} \delta^{n-1}(X) :\bar{\phi}(x_{n}) \phi(x_{n}): 
\partial^{\mu}A_{\mu}(x_{n}) u(x_{n}).
\ee

But we have:
\bea
{\cal P}'_{10} = d_{Q} \left[ -{i\over 2} c_{1} \delta^{n-1}(X) 
:\bar{\phi}(x_{n}) \phi(x_{n}): A_{\mu}(x_{n}) A^{\mu}(x_{n}) \right],
\nonumber \\
{\cal P}'_{11} = d_{Q} \left[ i c_{2} \delta^{n-1}(X) 
:\bar{\phi}(x_{n}) \phi(x_{n}): \tilde{u}(x_{n}) u(x_{n}): \right]
\eea
so we can make the redefinition (\ref{redef2}) and eliminate these terms.

14') In this case the generic form of the numerical distribution is again
(\ref{p2}) and we obtain terms of the type studied at 9').

The proof is finished.
$\qed$


\section{Conclusions}

We have succeed to give complete proof of the renormalizability of the
quantum electrodynamics. It is much simpler that the proofs from the
literature based on the usual BRST transformation (see \cite{We} and literature
quoted there). Its main advantages, beside the conceptual clearness, are:
\begin{enumerate}
\item
the r\^ole of Feynman graphs is minimal (only in writing Wick theorem);
\item
we circumvent the problem of the so-called 1-particle reducible graphs (see the
end of the preceding Subsection.
\end{enumerate}

It is an interesting problem to extend this analysis to the case of the
standard model. This was done in a series of papers \cite{DHKS1}, \cite{DHKS2},
\cite{DHS3},\cite{DHS3} using beside Epstein-Glaser approach some rather
complicated identities (the so-called C-g identities) and group theoretical
analysis. It is to be expected that our method could simplify somewhat the
analysis of this model and of other gauge models of physical interest from the
literature. A step in this direction is made in \cite{D1} where the case of
Yang-Mills fields of zero mass coupled with Dirac fields through vectorial
currents is investigated. We propose to extend this analysis to the general
case where scalar ghosts and axial currents are present in future papers.

It is natural to expect that this approach to renormalization theory gives the
same result as the usual procedure of Becchi, Rouet, Stora and Tyutin. A proof
of this fact based on the quantum Noether method \cite{HS} appears in
\cite{BHHS}. However, we can give here a much simpler argument. It is clear
that both approaches verify the same set of axioms so one can use the
characterization of the non-uniqueness given above.

\end{document}